%% file: GraphAnomaly.tex
\RequirePackage{amsmath}	

\documentclass[twocolumn]{svjour3}

\usepackage[T1]{fontenc}
\usepackage[utf8]{inputenc}

\usepackage{newtxtext,newtxmath}

\usepackage{graphicx}
\usepackage{bm,color,url,multirow,booktabs,amssymb}
\usepackage[caption=false]{subfig} 
\usepackage[acronym]{glossaries}
\usepackage{hyperref}
\usepackage[utf8]{inputenc}

\usepackage{pgfplots}
\pgfplotsset{compat=1.5}

\newcommand{\term}[1]{\emph{#1}}
\newcommand{\scanname}[1]{\texttt{#1}}
\newcommand{\figref}[1]{\hyperref[#1]{Figure~\ref*{#1}}}
\newcommand{\tabref}[1]{\hyperref[#1]{Table~\ref*{#1}}}
\newcommand{\secref}[1]{\hyperref[#1]{Section~\ref*{#1}}}
\newcommand{\vc}[1]{{\bf #1}}
\newcommand{\x}[1]{\vc{x}_{#1}}

\newcommand{\ox}[1]{\overline{\vc{x}}_{#1}}

\newcommand{\s}[1]{\vc{s}_{#1}}
\newcommand{\ts}[1]{\widetilde{\vc{s}}_{#1}}

\newcommand{\y}[1]{\vc{y}_{#1}}
\def\vmu{\hat{\bm \mu}}
\def\vC{\widehat{\vc{C}}}
\def\vQ{\widehat{\vc{Q}}}

\def\vLn{\vc{L}_{Q}}
\def\vLc{\vc{L}_{C}}
\def\ie{\emph{i.e.,}}
\def\eg{\emph{e.g.,}}

\def\rocsize{5.5cm}
\def\rocsizeT{5.5cm}
\def\energysize{.3\linewidth}
\def\resultsize{.16\linewidth}
\def\tumor{\emph{Tumor}}
\def\real{\emph{Field}}
\def\impl{\emph{Impl-14}}
\def\implB{\emph{Impl-4}}
\def\urbA{\emph{Urban-A}}
\def\urbB{\emph{Urban-B}}

\newacronym{pet}{PET}{positron emission tomography}
\newacronym{pca}{PCA}{principal component analysis}
\newacronym{id}{ID}{intrinsic dimensionality}
\newacronym{dyn}{dyn-PET}{dynamic PET}
\newacronym{fdg}{FDG-PET}{fluorodeoxyglucose positron emission tomography}
\newacronym{ircc}{IRCCS-FPO}{Candiolo Cancer Institute}
\newacronym{tac}{TAC}{time-activity curve}
\newacronym{rxd}{RXD}{Reed--Xiaoli detector}
\newacronym{wscf}{WSCF}{whitening spatial correlation filtering}
\newacronym{rsad}{RSAD}{Random-selection-based anomaly detector}
\newacronym{gft}{GFT}{graph Fourier transform}
\newacronym{dct}{DCT}{discrete cosine transform}
\newacronym[longplural={regions of interest}]{roi}{ROI}{region of interest}
\newacronym{soi}{SOI}{spatial overlap index}
\newacronym{roc}{ROC}{receiver operating characteristic}
\newacronym{fpr}{FPR}{false positive rate}
\newacronym{gmrf}{GMRF}{gaussian Markov random field}
\newacronym{klt}{KLT}{Karhunen--Lo\`{e}ve transform}
\newacronym{our}{LAD}{Laplacian anomaly detector}
\newacronym{auc}{AUC}{area under the curve}

\begin{document}

\title{Graph Laplacian for Image Anomaly Detection}

\author{Francesco~Verdoja and Marco~Grangetto}

\institute{F. Verdoja \at
	Aalto University, School of Electrical Engineering \\
	Maarintie 8, Espoo, Finland \\
	\email{francesco.verdoja@aalto.fi}           
	\and
	M. Grangetto \at
	University of Turin, Department of Computer Science \\
	Via Pessinetto 12, Turin, Italy \\
	\email{marco.grangetto@unito.it}           
}

\date{}

\maketitle

\begin{abstract}
\gls{rxd} is recognized as the benchmark algorithm for image anomaly 
detection; however, it presents known limitations, namely the dependence over 
the image following a multivariate Gaussian model, the estimation and inversion 
of a high-dimensional covariance matrix, and the inability to effectively 
include spatial awareness in its evaluation. In this work, a novel graph-based 
solution to the image anomaly detection problem is proposed; leveraging the
graph Fourier transform, we are able to overcome some of \gls{rxd}'s 
limitations while reducing computational cost at the same time. Tests over both 
hyperspectral and medical images, using both synthetic and real anomalies, 
prove the proposed technique is able to obtain significant gains over 
performance by other algorithms in the state of the art.
\end{abstract}

\glsreset{rxd}


\input{sections/1_intro}

\input{sections/2_back}

\input{sections/3_related}

\input{sections/4_tech}

\input{sections/5_exp}

\input{sections/6_disc}

\input{sections/7_concl}

\bibliographystyle{spmpsci} 
\bibliography{refs}

\end{document}

%% file: sections/1_intro.tex
\section{Introduction} \label{sec:intro}

Anomaly detection is the task of spotting items that do not conform to 
the expected pattern of the data. In the case of images, it usually refers to 
the problem of spotting pixels showing a peculiar spectral signature when 
compared to all other pixels in an image. Image anomaly detection is considered 
one of the most interesting and crucial tasks for many high-level image- and 
video-based applications, \eg{} surveillance, environmental monitoring, and 
medical analysis~\cite{cheng_gaussian_2015}.

One of the most 
used and widely validated techniques for anomaly detection is 
Reed-Xiaoli detector, often called RX detector for 
short~\cite{reed_adaptive_1990}, which is the most known example of 
covariance-based anomaly detectors.
This class of detectors has found wide adoption in many domains, from 
hyperspectral~\cite{matteoli_tutorial_2010} to medical 
images~\cite{verdoja_automatic_2014}; however, methods of this type suffer from 
crucial drawbacks, most noticeably the need for covariance estimation and 
inversion. Many situations exist where the drawbacks of these state-of-the-art 
anomaly detectors lead to poor and unreliable results~\cite{yuan_fast_2015}. 
Moreover, the operations required by those techniques are computationally expensive \cite{chang_anomaly_2002}.
For all these reasons, the research for a fast and reliable image anomaly detection
strategy able to overcome the limitations of covariance-based anomaly detectors 
deserves further efforts.

In this paper, we use graphs to tackle image anomaly detection.
Graphs are proved to be natural tools to represent data in many 
domains, \eg{} recommendation systems, social networks, or protein interaction 
systems~\cite{chung_spectral_1997}. Recently, they have found wide adoption 
also in computer vision and image processing communities, thanks to their 
ability to intuitively model relations between pixels. Graph-based approaches 
have been proposed to this date to solve a wide variety of image processing 
tasks, \eg{} edge detection~\cite{baterina_image_2010}, gradient 
estimation~\cite{ravazzi_curl-constrained_2017}, and 
segmentation~\cite{boykov_graph_2006,santner_interactive_2011}.
In particular, spectral graph theory has been recently bridged with signal 
processing, where the graph is used to model local relations between signal 
samples~\cite{sandryhaila_discrete_2013,shuman_emerging_2013}. As an example, 
graph-based signal processing is emerging as a novel approach in the design of 
energy compacting image 
transformations~\cite{fracastoro_steerable_2017,fracastoro_predictive_2015,hu_multiresolution_2015,verdoja_directional_2017,zhang_analyzing_2013}.

To this date, graph-based approaches have not been proposed for image anomaly 
detection, although many techniques for anomaly detection on generic graphs 
have been explored in the literature~\cite{akoglu_graph_2014}.
Those techniques cannot be straightforwardly extended to images since they 
usually exploit anomalies in the topology of the graph to extract knowledge 
about the data~\cite{chung_spectral_1997}.
On the other hand, in the image case the graph topology is constrained to the 
pixel grid, whereas different weights are assigned to edges connecting pixels 
depending on their similarity or correlation.

Our proposed approach uses an undirected weighted graph to model the 
expected behavior of the data and then computes the distance of each pixel in 
the image from the model. We propose to use a graph to model  
spectral or both spectral and spatial correlation. The main 
contribution of this paper is a novel anomaly detection approach which exploits 
spectral graph theory to overcome one of the well-known 
limitations of RX detector and other covariance-based anomaly detectors, \ie{} 
the need to estimate and invert a covariance matrix. Estimation of the 
covariance may be very critical in the presence of a small sample size; moreover, 
inverting such a matrix is also a complex, badly conditioned and unstable 
operation~\cite{khazai_anomaly_2011}. Our novel anomaly detector estimates the 
statistic of the background using a graph Laplacian matrix. Also, the graph 
model used by our approach is abstract and flexible enough to be tailored to 
any prior knowledge of the data possibly available. 
The effectiveness of our methodological contributions is shown in two 
use cases: a typical hyperspectral anomaly detection experiment and a novel 
application for tumor detection in 3D biomedical images.

The paper is organized as follows: we will first give a brief overview of RX 
detector and the graph Fourier transform in \secref{sec:back} and go over some 
related work in \secref{sec:related}, and then we will present our technique in 
\secref{sec:tech}; we will then evaluate the performance of our technique and 
compare our results with those yielded by algorithms in the state of the art 
both visually and objectively in \secref{sec:exp}, and we will discuss these 
results in \secref{sec:disc}; finally, conclusions will be drawn in 
\secref{sec:concl}.

%% file: sections/2_back.tex
\section{Background} \label{sec:back}

Anomaly detection refers to a particular class of target detection problems, 
namely the ones where no prior information about the target is available. In 
this scenario, supervised approaches that try to find pixels which match 
reference spectral characteristics 
(\eg{}~\cite{du_real-time_2003,kwon_kernel_2006}) cannot usually be employed.
This extends also to supervised deep learning 
or other data-driven approaches, which attempt to learn a parametric model
from a set of labeled data. Although deep learning methods have
found increasingly wide adoption for many other tasks in image processing and 
computer vision \cite{cheng_learning_2016,han_representing_2013,zhang_co-saliency_2017}, 
their application to anomaly detection---especially on hyperspectral 
and medical imaging---is stifled by multiple factors: first, pixels have to
be considered anomalous according to intra-image 
metrics which are difficult to capture in a dataset; second, the amount of data required
to train the models is not often available in these contexts
\cite{chalapathy_deep_2019,li_transferred_2017}.
For these reasons, classical unsupervised approaches are preferable instead. These
algorithms detect anomalous or peculiar pixels showing high spectral distance
from their surrounding~\cite{du_random-selection-based_2011}. To this end, the
typical strategy is to extract knowledge of the background statistics from the
data and then measure the deviation of each examined pixel from the learned 
knowledge according to some affinity function.

\subsection{\acrlong{rxd}} \label{sec:rx}

The best known and most widely employed algorithm for anomaly detection is 
\gls{rxd} by Reed and Yu~\cite{reed_adaptive_1990}. To this date, it is 
still used as a benchmark algorithm for many anomaly detection 
applications~\cite{banerjee_support_2006,du_random-selection-based_2011,manolakis_is_2009,matteoli_models_2013}.
\gls{rxd} assumes the background to be characterized by a non-stationary 
multivariate Gaussian model, estimated by the image mean and covariance. Then, 
it measures the squared Mahalanobis 
distance~\cite{mahalanobis_generalized_1936} of each pixel from the estimated 
background model. Pixels showing distance values over a set threshold are 
assessed to be anomalous.

Formally, \gls{rxd} works as follows.
Consider an image $\vc{I} = [\x{1}\x{2}\ldots\x{N}]$ consisting of~$N$ 
pixels, where the column vector $\x{i} = [x_{i1} x_{i2} \ldots x_{im}]^T$ 
represents the value of the $i$-th pixel over the~$m$ channels (or spectral 
bands) of~$\vc{I}$.
The expected behavior of background pixels can be captured by the 
mean vector~$\vmu$ and covariance matrix~$\vC$ which are estimated as follows:
\begin{equation} \label{eq:cov}
\vmu = \frac {1}{N} \sum_{i=1}^N \x{i} \enspace\mbox{,~~and~~} \vC = \frac 
{1}{N} \sum_{i=1}^N \ox{i}\ox{i}^T\enspace,
\end{equation}
where $\ox{i} = (\x{i} - \vmu)$.

Mean vector and covariance matrix are computed under the assumption that 
vectors~$\x{i}$ are observations of the same random process; it is usually 
possible to make this assumption as the anomaly is small enough to have a
negligible impact on the estimate~\cite{chang_anomaly_2002}.

Then, the generalized likelihood of a pixel~$\x{}$ to be anomalous with respect 
to the model~$\vC$ is expressed in terms of the square of the Mahalanobis 
distance~\cite{mahalanobis_generalized_1936}, as follows:
\begin{equation} \label{eq:detector}
\delta_{\gls{rxd}}(\x{}) \; = \; \ox{}^T \: 
\vQ \: \ox{}\enspace,
\end{equation}
where $\vQ = \vC^{-1}$, \ie{} the inverse of the covariance matrix, also known 
in the literature as the precision matrix.

Finally, a decision threshold~$\eta$ is usually employed to confirm or refuse 
the anomaly hypothesis. A common approach is to set~$\eta$ adaptively as a 
percentage of~$\delta_{\gls{rxd}}$ dynamic range as follows:
\begin{equation} \label{eq:eta}
\eta = t \cdotp \max_{i = 1,\dots, N}(\delta_{\gls{rxd}}(\x{i}))\enspace,
\end{equation}
with $t \in [0, 1]$. Then, if $\delta_{\gls{rxd}}(\x{}) \geq \eta$, the 
pixel~$\x{}$ is considered anomalous.

An interesting property of \gls{rxd} has been observed by Chang and Heinz
in~\cite{chang_constrained_2000}. In that work, the authors demonstrated 
how \gls{rxd} can be considered an inverse operation of the \gls{pca}.

More precisely, let us assume that $\kappa_1 \geq \kappa_2 \geq \ldots \geq 
\kappa_m$ are the eigenvalues of the $m \times m$ covariance matrix~$\vC$ and 
$\{\vc{v}_1, \vc{v}_2, \ldots, \vc{v}_m\}$ is its 
set of unit eigenvectors with~$\vc{v}_j$ corresponding to~$\kappa_j$. We can 
then form the matrix $\vc{V} = [\vc{v}_1 \vc{v}_2 \ldots \vc{v}_m]$ with 
the $j$-th column 
specified by~$\vc{v}_j$. $\vc{V}$ can be used to decorrelate the signal by 
diagonalizing~$\vC$ into the 
diagonal matrix~$\vc{K}$ whose $j$-th diagonal element is~$\kappa_j$, 
such that $\vc{V}^T \vC \vc{V} = \vc{K}$ and $\vc{V}^T \vQ \vc{V} = 
\vc{K}^{-1}$. Then, we can compute $\y{} = \vc{V}^T \ox{}$, which is 
known as the \gls{klt}. Data dimensionality reduction via \gls{pca} usually 
involves the computation of~$\y{}$ using just the first $p \ll m$ columns 
of~$\vc{V}$.
As shown in~\cite{chang_constrained_2000}, \eqref{eq:detector}  can be 
expressed as a function of~$\y{}$ as follows:
\begin{equation} \label{eq:ipca}
   \begin{array}{rcl}
      \delta_{\gls{rxd}}(\x{}) &=& \ox{}^T \: \vQ \: \ox{}\\[.4em]
      &=& (\vc{V}\y{})^T \: \vQ \: (\vc{V}\y{})\\[.4em]
      &=& \y{}^T \: (\vc{V}^T \vQ \vc{V}) \: \y{}\\[.4em]
      &=& \y{}^T \vc{K}^{-1} \y{}\\[.4em]
      &=& \sum_{j=1}^{m} \kappa_j^{-1} y_{j}^2\enspace,
   \end{array}
\end{equation}
where~$y_j$ represents the $j$-th element of the KLT vector~$\y{}$.

From this formulation, one can notice that \gls{rxd} detects targets 
with small energies that are represented
by small eigenvalues. This is because, according to \eqref{eq:ipca}, the 
smaller the eigenvalue, the greater its contribution to the value 
of~$\delta_{\gls{rxd}}$. This is reasonable, since if an anomalous small 
target is present in the image, it will not be visible in the principal 
components, but it is rather going to appear in smaller 
components~\cite{chang_anomaly_2002}. However, when seeing 
\gls{rxd} in this form, it is quite evident that the last components, which are 
those containing mostly noise, are actually weighted the most. To improve the 
result of \gls{rxd}, a value $p \ll m$ can be 
determined~\cite{harsanyi_determining_1993}. 
Then, the eigenvalues beyond the first (greater)~$p$ will be considered to 
represent components containing only noise and will be discarded.
We then obtain a de-noised version of \gls{rxd} that can be expressed as follows:
\begin{equation} \label{eq:rxp}
   \delta^p_{\gls{rxd}}(\x{}) = \sum_{j=1}^{p} \kappa_j^{-1} y_{j}^2\enspace.
\end{equation}
Obviously, $\delta^m_{\gls{rxd}} = \delta_{\gls{rxd}}$.

The issue of determining~$p$ was addressed 
in~\cite{chang_noise_1999,harsanyi_determining_1993} and is closely related to 
the problem of determining the \gls{id} of the image signal. Empirically, $p$ 
is usually set such that a desired percentage $\psi \in [0,1]$ of the original 
image cumulative energy content is retained. The cumulative energy content of 
the first~$p$ principal components of an image $\vc{I} = 
[\x{1}\x{2}\ldots\x{N}]$ can be expressed in terms of the image's \gls{klt} transform 
$\vc{Y} = \vc{V}^T\overline{\vc{I}} = [\y{1}\y{2}\ldots\y{N}]$ where 
$\overline{\vc{I}} = [\ox{1}\ox{2}\ldots\ox{N}]$ as
\begin{equation} \label{eq:energy}
 e(\vc{I},p) = \sum_{i=1}^{N}\sum_{j=1}^{p} y^2_{ij}\enspace,
\end{equation}
where~$y_{ij}$ is the $j$-th element of the vector~$\y{i}$.
We then choose the smallest $p \in [1,m]$, such that $e(\vc{I},p)/e(\vc{I},m) 
\leq \psi$. 
Commonly for dimensionality reduction applications $\psi = 0.9$, but for 
anomaly detection purposes that value might be too low, given we do not want to 
risk to lose the anomaly. In this case, $\psi = 0.99$ is usually more 
appropriate.
 
\subsection{Graph Fourier transform} \label{sec:gft}

In recent years, the growing interest in graph-based signal 
processing~\cite{sandryhaila_discrete_2014}
has stimulated the study of graph-based transform approaches. 
These methodologies map the image content onto a topological graph where nodes 
represent pixel intensities and edges model relations between nodes, \eg{} 
according to a criterion based on correlation or other similarity measures. 
The Fourier transform can be generalized to graphs obtaining the so-called 
\gls{gft}~\cite{sandryhaila_discrete_2013}. 

Consider an undirected, weighted graph $\mathcal{G} = (\mathcal{V}, 
\mathcal{E})$ composed of a vertex set~$\mathcal{V}$ of order~$n$ and an edge 
set~$\mathcal{E}$ specified by $(a, b, w_{ab})$, where $a, b \in \mathcal{V}$, 
and $w_{ab} \in \mathbb{R}^+$ is the edge weight between vertices~$a$ and~$b$. 
Thus, a weighted graph can be described by its adjacency matrix~$\vc{W}$ where 
$\vc{W}(a,b) = w_{ab}$. A graph signal is a mapping that assigns a value to 
each vertex, denoted as $\s{} = [s_1 s_2\ldots s_n]^T$.

Typically, when computing the \gls{gft} a graph is constructed to capture
the inter-pixel correlation and is used to compute the optimal 
decorrelating transform leveraging on spectral graph 
theory~\cite{shuman_emerging_2013}.
From the adjacency (also called weight) matrix~$\vc{W}$, the 
\term{combinatorial graph Laplacian} matrix $\vc{L} = \vc{D}-\vc{W}$ can be 
computed, where~$\vc{D}$ is the degree matrix: 
a diagonal matrix whose $a$-th diagonal element is equal to the sum of 
the weights of all edges incident to the node~$a$. Formally,
\begin{equation} \label{eq:degree}
\vc{D}(a,b) = \begin{cases}
   \sum^n_{k=1} w_{ak} & \mbox{if } a = b,\\
   0 & \mbox{otherwise.}
\end{cases}
\end{equation}

In some scenarios, it is useful to normalize weights in the Laplacian matrix; in those 
cases, the use of the \term{symmetric normalized Laplacian 
matrix}~$\vc{L}^{sym}$ is preferred. It is defined as 
\begin{equation} \label{eq:lsym}
\vc{L}^{sym} = \vc{D}^{-\frac{1}{2}} \vc{L} \vc{D}^{-\frac{1}{2}}\enspace.
\end{equation}
$\vc{L}^{sym}$ has important properties, \ie{} its eigenvalues are 
always real, nonnegative, and bounded into the range $[0,2]$; for these reasons, 
the spectrum of a symmetric normalized Laplacian relates well to other graph 
invariants for general graphs in a way that other definitions fail to 
do~\cite{chung_spectral_1997}.

Any Laplacian matrix~$\vc{L}$ is a symmetric positive semi-definitive matrix 
with eigendecomposition:
\begin{equation} \label{eq:laplac}
\vc{L}=\vc{U} \vc{\Lambda} \vc{U}^T\enspace,
\end{equation}
where~$\vc{U}$ is the matrix whose columns are the eigenvectors of~$\vc{L}$ 
and~$\vc{\Lambda}$ is the diagonal matrix whose diagonal elements are the 
corresponding eigenvalues.
The matrix~$\vc{U}$ is used to compute the \gls{gft} of a signal~$\s{}$ 
as:
\begin{equation} \label{eq:gft}
\ts{} =\vc{U}^T \s{}\enspace.
\end{equation}

The inverse \gls{gft} is then given by
\begin{equation} \label{eq:igft}
\s{}=\vc{U} \ts{}\enspace.
\end{equation}

When computing the \gls{gft}, the eigenvalues in~$\vc{\Lambda}$ are usually 
sorted for increasing magnitude, the first eigenvalue being equal to 
zero~\cite{sandryhaila_discrete_2013}, \ie{} $0 = \lambda_1 \leq \lambda_2 \leq 
\ldots \leq \lambda_m$. The eigenvectors in~$\vc{U}$ are sorted accordingly.

%% file: sections/3_related.tex
\section{Related work} \label{sec:related}

Despite its popularity, \gls{rxd} has recognized drawbacks that 
undermine its 
performance in some applications. For a full discussion over the limitations of 
\gls{rxd}, we suggest~\cite{chang_anomaly_2002,yuan_fast_2015}; however, they can 
be summarized in the following:
\begin{enumerate}
	\item \gls{rxd} involves a high-dimensional covariance matrix 
	that needs to be estimated and inverted, often under a small sample 
	size~\cite{banerjee_support_2006,khazai_anomaly_2011}. Those are unstable, 
	highly complex, and badly conditioned operations;
	\item \gls{rxd} often suffers from high 
	\gls{fpr}~\cite{banerjee_support_2006,gurram_support_2011,matteoli_models_2013};
	\item \gls{rxd} assumes that the background follows a multivariate Gaussian 
	model, but there are cases in which this assumption might not be adequate, 
	\eg{} in the case of multiple materials and 
	textures~\cite{banerjee_support_2006,chang_anomaly_2002,du_discriminative_2014,gurram_support_2011};
	\item \gls{rxd} lacks spatial awareness: every pixel is evaluated 
	individually extrapolated from its context~\cite{gaucel_whitening_2005}.
\end{enumerate}

To address these issues, recent works have iterated over \gls{rxd}'s 
idea, \eg{} 
by considering subspace features~\cite{du_target_2014,stein_anomaly_2002}, by 
using kernels to go beyond the Gaussian 
assumption~\cite{du_discriminative_2014,kwon_kernel_2005}, by applying 
dimensionality reduction~\cite{gu_selective_2008}, by improving how the 
background statistics are 
estimated~\cite{du_random-selection-based_2011,matteoli_hyperspectral_2011}, or 
by exploiting sparsity and compress sensing 
theory~\cite{du_beyond_2016,fowler_anomaly_2012,zhang_sparse_2015}.
In this work, we generalize \gls{rxd}'s idea by looking at it from the point of 
view of spectral graph theory. This not only makes us able to avoid costly 
covariance matrix inversions, but also allows us to incorporate spatial 
information and any prior knowledge about the background model into the 
detector. Previous work trying to including spatial awareness in the detector 
is available in the literature, a noteworthy example is \gls{wscf} 
\cite{gaucel_whitening_2005}, where the authors propose to apply a whitening 
transformation based on the eigendecomposition of the image covariance matrix. 
On the whitened space, \gls{rxd} is represented by the 
Euclidean norm. Then, by using an approach based on constrained energy 
minimization, \gls{wscf} spots anomalous pixels by estimating consistency to 
their neighborhood in the whitened space. We compare our proposed approach to 
\gls{wscf} in the experimental section.

Although prior research targeting anomaly detection in graphs exists, 
it mostly 
focuses on anomalies in a graph structure, and not on graph 
signals~\cite{akoglu_graph_2014,chung_spectral_1997}. For example, in the 
context of behavioral monitoring and intelligence, the structure of social 
graphs can be analyzed to spot subgraphs expressing patterns deviating from 
the rest of the network~\cite{noble_graph-based_2003}. However, in images, the 
structure of the graph is fixed to a grid, and the application of graph-based 
anomaly detection algorithms coming from other domains is not straightforward; 
even in works where peculiarities in the graph signal are under observation, 
structure is included as part of the signal, as for example 
in~\cite{egilmez_spectral_2014} where a signal function of the physical 
distance between wireless sensors is proposed. The effectiveness of these 
approaches to images has not been reported yet.

Our proposed graph-based approach is founded on two recent findings: 
first, Zhang and Flor\^{e}ncio~\cite{zhang_analyzing_2013} have shown that a 
Laplacian model can be used as an estimation of the precision matrix~$\vc{Q}$ 
of an image, under the assumption that the image follows a \gls{gmrf} model. This 
amounts to using a function of the partial correlation between nodes as graph 
weights.
Second, it has been demonstrated how the \gls{gft} can be considered an 
approximation of the \gls{klt} for graph signals 
\cite{hu_multiresolution_2015}. Recent literature in spectral graph theory has 
exploited this relationship to provide 
novel graph-based solutions to classical signal processing problems, in 
particular for image compression where the use of the \gls{gft} has been 
proposed as an alternative to the \gls{dct} 
\cite{cheung_graph_2018,fracastoro_steerable_2017,fracastoro_predictive_2015,hu_multiresolution_2015}.
This relationship is, however, never been explored in the context of image 
anomaly detection, which motivated us to study it in this work.

%% file: sections/4_tech.tex
\section{Method} \label{sec:tech}

In this work, we exploit the analogy between \gls{klt} and \gls{gft} in the 
framework of anomaly detection. In the \gls{gft} definition, the role of the 
covariance matrix in the \gls{klt} is taken by the graph Laplacian. 
It turns out that~$\vc{L}$ can be exploited also in the inverse problem of 
anomaly detection according to \eqref{eq:ipca}. 
We here propose a novel algorithm for image anomaly detection, which we will 
refer to as \gls{our}. \gls{our} overcomes some of the known limitations of 
\gls{rxd} exposed in \secref{sec:rx}: it can be used to avoid problematic 
covariance matrix estimate and inversion, and it is able to include spatial 
information as well as a priori knowledge, when available.

\subsection{Construction of the graph model} \label{sec:model}

Given an image~$\vc{I}$ composed of~$N$ pixels and having~$m$ spectral bands or 
channels, we first build an undirected graph $\mathcal{G} = (\mathcal{V}, 
\mathcal{E})$ to serve as the model for the background pixels in the image. The 
graph is used to model local relations between pixel values and can be 
constructed to capture spectral and spatial characteristics. 
Topology and weights of the graph have to be chosen accordingly to the 
domain. We will discuss some general construction strategies in 
\secref{sec:spectr} and \secref{sec:space}. The chosen graph will be described 
by a weight matrix~$\vc{W}$, from which a Laplacian matrix~$\vc{L}$ will be 
computed according to the procedure detailed in \secref{sec:gft}. The use of 
the symmetric normalized Laplacian, constructed as in \eqref{eq:lsym}, in place 
of the unnormalized combinatorial one is to be preferred for the reasons 
expressed in \secref{sec:gft}. Also, $\vc{L}^{sym}$ is proved to be preferable in 
similar domains, \eg{} segmentation and 
classification~\cite{bertozzi_diffuse_2012,galasso_spectral_2014}.

\subsection{Graph-based anomaly detection} \label{sec:detection}

Given a pixel~$\x{}$, we define a corresponding graph signal~$\s{}$, \eg{} 
describing the spectral bands of~$\x{}$ or its spatial neighborhood, and 
compute the distance of~$\x{}$ from the model as
\begin{equation} \label{eq:lad}
\begin{array}{rcl}
\delta_{\gls{our}}(\x{}) &=& \s{}^T \: \vc{L} \: \s{}\\[.4em]
   &=& (\vc{U}\ts{})^T \: \vc{L} \: (\vc{U}\ts{})\\[.4em]
&=& \ts{}^T \: (\vc{U}^T \vc{L} \vc{U}) \: \ts{}\\[.4em]
&=& \ts{}^T \: \vc{\Lambda} \: \ts{}\\[.4em]
&=& \sum_{j=1}^{m} \lambda_j \: \widetilde{s}_{j}^2\enspace,
\end{array}
\end{equation}
where~$\widetilde{s}_j$ represents the $j$-th element of the \gls{gft} 
vector~$\ts{}$, and~$\vc{U}$ and~$\vc{\Lambda}$ refer to the eigenvector and 
eigenvalue matrices used for the eigendecomposition of~$\vc{L}$ in 
\eqref{eq:laplac}.
Although this formulation might look similar to the one of \gls{rxd} given in 
\eqref{eq:ipca}, some important differences have to be noted. First, the model 
used is not the inverse of the covariance matrix~$\vC^{-1}$, but an arbitrary 
Laplacian model; this is a generalization over \gls{rxd}, because if the image 
follows a \gls{gmrf} model, then a Laplacian can be constructed to estimate the 
precision matrix~\cite{zhang_analyzing_2013}, but if this is not the case a 
Laplacian model can be computed according to any knowledge of the domain. 
Second, the Laplacian matrix can be used to capture both spatial and spectral 
characteristics as we will detail in \secref{sec:space}. Another thing to 
notice is that in \eqref{eq:lad} each contribution~$\widetilde{s}_j$ is 
multiplied by~$\lambda_j$, whereas in \gls{rxd} each~$y_j$ was instead divided 
by the corresponding eigenvalue~$\kappa_j$. 

As already discussed for \gls{rxd}, we can also use a de-noised version of the 
\gls{gft} where only the first smaller $p \ll m$ eigenvectors are kept, 
removing the higher and noisier frequencies and obtaining the following:
\begin{equation} \label{eq:ladp}
\delta^p_{\gls{our}}(\x{}) = \sum_{j=1}^{p} \lambda_j 
\widetilde{s}_{j}^2\enspace.
\end{equation}
The parameter~$p$ is determined accordingly to the percentage of retained 
cumulative energy, following the approach presented in \secref{sec:rx}.

Finally, a decision threshold over~$\delta_{\gls{our}}$ is needed to determine 
if a pixel is anomalous or not. An approach similar to the one described in 
\secref{sec:rx} can be employed.

\begin{figure}
   \centering
   \subfloat[Spectral 
   connectivity]{\includegraphics[width=0.49\linewidth]{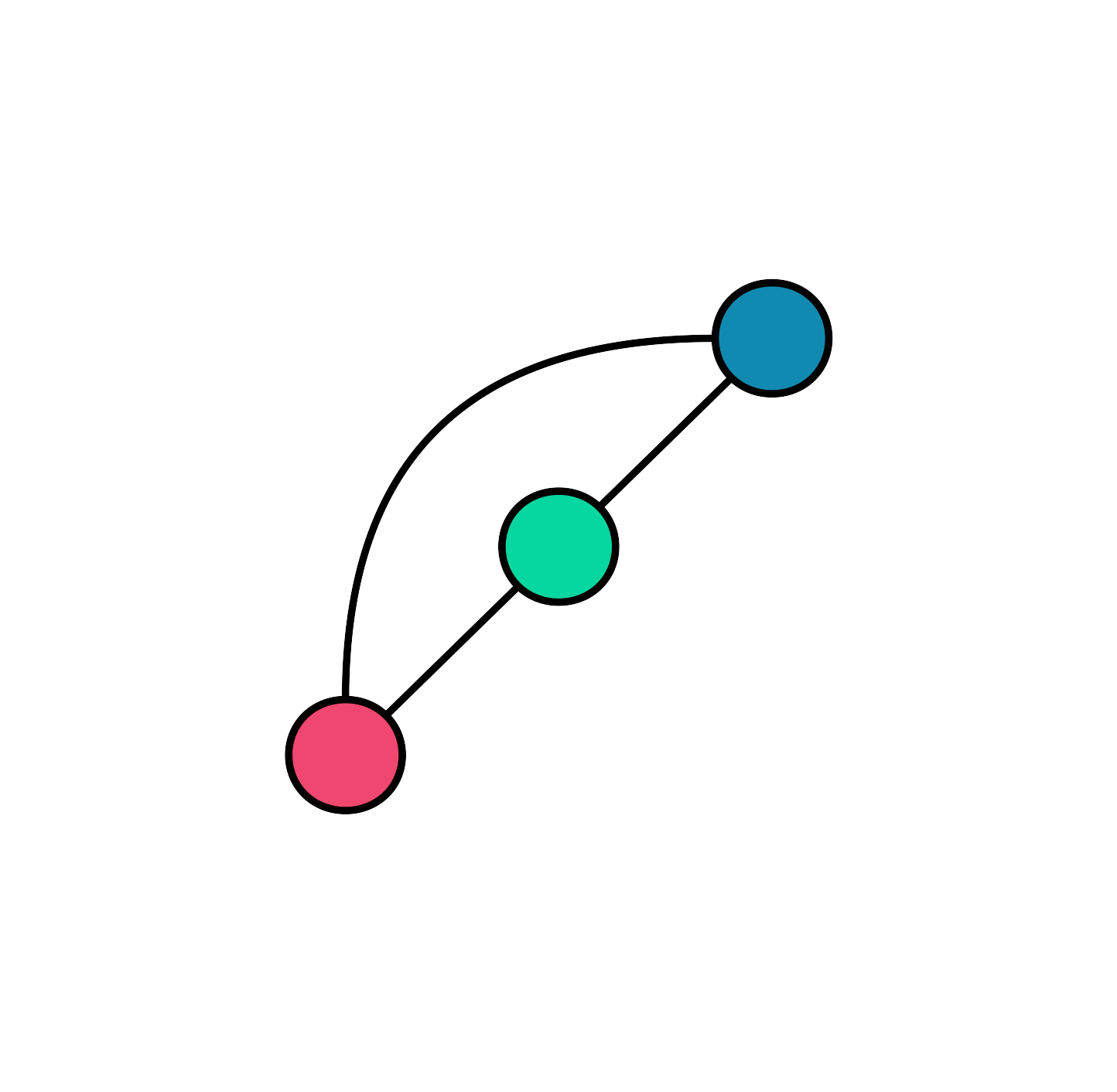}\label{fig:spectr}}
   \hfil
   \subfloat[Spatial 
   connectivity]{\includegraphics[width=0.49\linewidth]{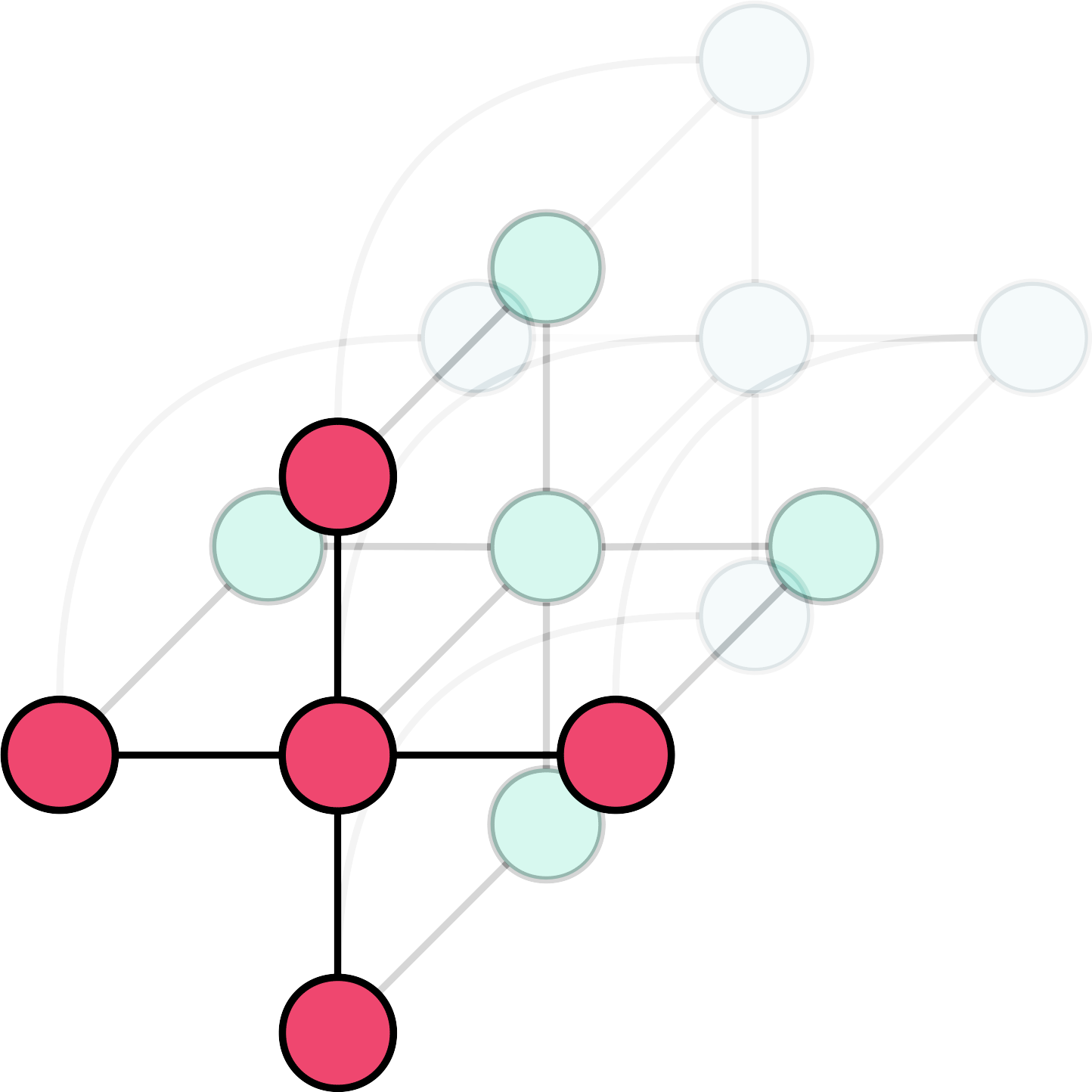}\label{fig:space}}
   \caption{\label{fig:graph}Example of 3-band graph connectivity: the spectral 
   components are fully connected, while spatially pixels are 4-connected.} 
\end{figure}

\subsection{Spectral graph model} \label{sec:spectr}

As already mentioned, the graph model is used to characterize the typical behavior around the pixel being tested for anomaly.
As in the case of standard \gls{rxd}, the graph can be employed to model only the 
spectral relations: in this case, the vertex set~$\mathcal{V}$ consists of~$m$ 
nodes, each representing one of the spectral bands of~$\vc{I}$; then, we connect each 
pair of nodes (bands) with an edge, obtaining a fully connected graph. An 
example of this topology for a 3-band image is shown in \figref{fig:spectr}. A 
weight is then assigned to each edge: 
if some a priori knowledge about inter-band correlation is available, it can be 
used to set weights 
accordingly; if this is not the case, a possibility is to use the image 
data to estimate the weights. Also, for each pixel~$\x{}$, the graph 
signal~$\s{}$ will contain exactly the value of that pixel over the~$m$ bands, 
after removing the mean; thus, $\s{} = \ox{}$.

Under the assumption that the image follows a \gls{gmrf} model, we might use  
partial correlation as weight, as proposed by Zhang and 
Flor\^{e}ncio~\cite{zhang_analyzing_2013}. To this end, given the precision 
matrix $\vQ = \vC^{-1}$, estimated according to \eqref{eq:cov}, we can 
set the weight of the edge connecting nodes~$a$ and~$b$ as:
\begin{equation} \label{eq:partial}
w_{ab} = - \frac{\vQ(a,b)}{\sqrt{\vQ(a,a) \: \vQ(b,b)}}\enspace.
\end{equation}
Note that $w_{aa} = 0$ as we do not include self-loops.
However, this approach still relies on the estimate and inversion of the 
covariance matrix that, as we already discussed, might be unreliable 
(especially in the presence of a small data sample) as well as expensive to 
compute: matrix inversion requires $O(m^3)$ time~\cite{lezoray_image_2012}.
Also, if the image does not follow a \gls{gmrf} model, this distance 
function might produce unreliable 
results, as for all other covariance-based methods. An option to safeguard against this could be to use the graph constructed to evaluate the \gls{gmrf} hypothesis
with an approach similar to the one proposed in \cite{anandkumar_detection_2009}

Another possibility is to use a different weight function, \eg{} the Cauchy function~\cite{grady_discrete_2010}, 
which has been proved to be able to capture graph distances effectively
for image signals and is commonly used as graph weight in other 
applications like image segmentation and compression~\cite{black_robust_1998,fracastoro_predictive_2015}.
We propose to set the weight of the edge connecting bands~$a$ and~$b$, according to the band 
mean vector $\vmu = [\mu_1 \mu_2 \ldots \mu_m]^T$ estimated as in 
\eqref{eq:cov}, as
\begin{equation} \label{eq:cauchy}
w_{ab} = \frac{1}{1 + \left(\frac{\mu_a - \mu_b}{\alpha}\right)^2}\enspace,
\end{equation}
where~$\alpha$ is a scaling parameter. In this study, we decided to set 
$\alpha = \frac{1}{m} \sum_{i=1}^m \mu_i$, to normalize all values according to 
the mean range of the bands. The advantages of this approach are twofold: 
it avoids using unreliable correlation estimates and does not require matrix 
inversion, thus reducing the computational cost significantly.

Although other approaches to estimate graph weights might be devised, in this 
study we will limit the analysis to these ones. 

\subsection{Integration of spatial information in the graph} \label{sec:space}

One of the advantages of using a graph-based approach is the flexibility of the 
model. For example, by augmenting the graph topology to include edges 
connecting each node to nodes describing the same band for the neighboring 
pixels, as shown in \figref{fig:space}, one is able to include spatial 
information in the model. 
We will refer to this spatially aware version of \gls{our} as \gls{our}-S.

When considering the case of 4-connected nodes, the resulting graph will be 
composed of $5m$ nodes; therefore, the weight matrix~$\vc{W}$, as well 
as the corresponding Laplacian matrix~$\vc{L}$, will be a $5m \times 5m$ 
matrix.
We can construct the weight matrix as follows:
\begin{equation} \label{eq:w-s}
\vc{W}(a,b) = \begin{cases}
   w'_{ab} & \mbox{if nodes $a, b$ represent different}\\[-0.2em]
   & \mbox{bands of the same pixel,}\\[0.2em]
   w''_{ab} & \mbox{if nodes $a, b$ belong to the same}\\[-0.2em]
   & \mbox{band of 4-connected pixels,}\\[0.2em]
   0 & \mbox{otherwise,}
\end{cases}
\end{equation}
where~$w'_{ab}$ and~$w''_{ab}$ are some spectral and spatial correlation 
measures, respectively.

Then, to compute the distance of a pixel~$\x{}$ from the model, a graph 
signal~$\s{}$ is constructed concatenating the vector corresponding to~$\x{}$ 
and its 4-connected neighbors; also in this case, the mean value~$\vmu$ 
is subtracted. It follows that the vector~$\s{}$ will have length $5m$.

The spectral weights~$w'_{ab}$ can be estimated as proposed in the previous 
section.
The weights~$w''_{ab}$ can be used to enforce a spatial prior: as an example in 
the following experimental analysis, we will set uniform spatial 
weights~$w''_{ab}=1$.

%% file: sections/5_exp.tex
\section{Experiments} \label{sec:exp}

To objectively evaluate \gls{our}'s performance, we selected a couple of 
scenarios in which the use of \gls{rxd} has been proposed. The first 
one is hyperspectral remote sensing, which is one of the most common use cases
for anomaly detection where the use of \gls{rxd} is widely 
validated~\cite{matteoli_tutorial_2010}; the second one is the domain of 3D 
volumetric segmentation of tumoral masses on \gls{pet} images, 
where we successfully explored the use of \gls{rxd} in the 
past~\cite{bracco_automatic_2016,verdoja_global_2016,verdoja_automatic_2014}. 
In these scenarios, we compare the performance of the proposed 
technique with 
those produced by \gls{rxd} and, in the hyperspectral domain, also with 
\gls{rsad}~\cite{du_random-selection-based_2011} and 
\gls{wscf}~\cite{gaucel_whitening_2005}. \Gls{rsad} employs multiple random 
selections of pixels to estimate the background statistics and then marks a 
pixel as anomalous by merging the output of the different runs by a majority 
voting approach. \Gls{wscf} applies a whitening 
transformation to the input based on the image covariance matrix and then 
incorporates spatial information in the anomaly measure. This latter algorithm 
is of particular interest for our evaluation, to compare its performance 
against our own spatially aware methodology.

\subsection{Hyperspectral remote sensing} \label{sec:hyper}

Hyperspectral images find wide adoption in remote sensing applications, where  
hyperspectral sensors are typically deployed on either aircraft
or satellites. The data produced by these sensors are a three-dimensional
array or ``cube'' of data with the width and length of the array corresponding
to spatial dimensions and the spectrum of each point as the third dimension.

\begin{figure}
	\centering
	\subfloat[Band 
	70]{\includegraphics[width=0.32\linewidth]{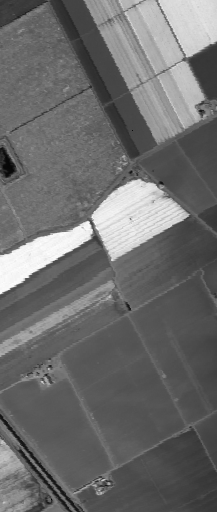}\label{fig:sal}}
	\hfil
	\subfloat[Classes]{\includegraphics[width=0.32\linewidth]{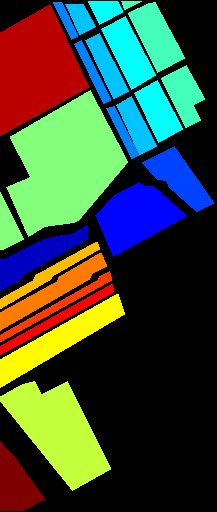}\label{fig:salc}}
	\caption{\label{fig:salinas}The full $512\times217$ Salinas scene}
\end{figure}

\begin{figure*}
	\centering
	\subfloat[Band 70 
	of 
	\urbA{}]{\includegraphics[width=\resultsize]{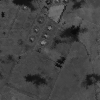}\label{fig:urb1crop}}
	\hfil
	\subfloat[Ground 
	truth]{\includegraphics[width=\resultsize]{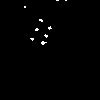}\label{fig:urb1gt}}
	\hfil
	\subfloat[\gls{rxd} ($t = 
	0.28$)]{\includegraphics[width=\resultsize]{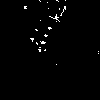}\label{fig:urb1rxt}}
	\hfil
	\subfloat[\gls{rsad}]{\includegraphics[width=\resultsize]{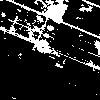}\label{fig:urb1rsadt}}
	\hfil
	\subfloat[\gls{wscf} ($t = 
	0.30$)]{\includegraphics[width=\resultsize]{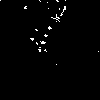}\label{fig:urb1wscft}}
	\hfil
	\subfloat[\gls{our} ($t = 
	0.36$)]{\includegraphics[width=\resultsize]{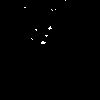}\label{fig:urb1ladt}}
	\hfil
	\subfloat[Band 20 
	of 
	\urbB{}]{\includegraphics[width=\resultsize]{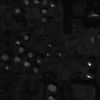}\label{fig:urb2crop}}
	\hfil
	\subfloat[Ground 
	truth]{\includegraphics[width=\resultsize]{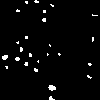}\label{fig:urb2gt}}
	\hfil
	\subfloat[\gls{rxd} ($t = 
	0.04$)]{\includegraphics[width=\resultsize]{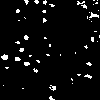}\label{fig:urb2rxt}}
	\hfil
	\subfloat[\gls{rsad}]{\includegraphics[width=\resultsize]{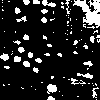}\label{fig:urb2rsadt}}
	\hfil
	\subfloat[\gls{wscf} ($t = 
	0.06$)]{\includegraphics[width=\resultsize]{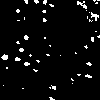}\label{fig:urb2wscft}}
	\hfil
	\subfloat[\gls{our} ($t = 
	0.02$)]{\includegraphics[width=\resultsize]{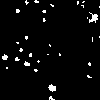}\label{fig:urb2ladt}}
	\hfil
	\subfloat[Band 70 
	of 
	\real{}]{\includegraphics[width=\resultsize]{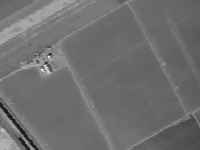}\label{fig:realcrop}}
	\hfil
	\subfloat[Ground 
	truth]{\includegraphics[width=\resultsize]{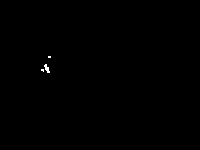}\label{fig:realgt}}
	\hfil
	\subfloat[\gls{rxd} ($t = 
	0.16$)]{\includegraphics[width=\resultsize]{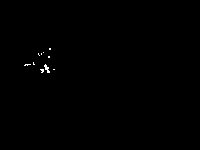}\label{fig:realrxt}}
	\hfil
	\subfloat[\gls{rsad}]{\includegraphics[width=\resultsize]{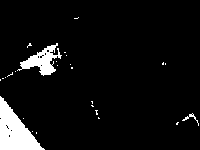}\label{fig:realrsadt}}
	\hfil
	\subfloat[\gls{wscf} ($t = 
	0.24$)]{\includegraphics[width=\resultsize]{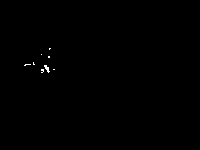}\label{fig:realwscft}}
	\hfil
	\subfloat[\gls{our} ($t = 
	0.46$)]{\includegraphics[width=\resultsize]{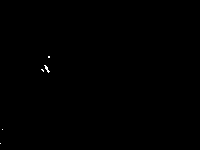}\label{fig:realladt}}
	\hfil
	\subfloat[Band 70 
	of 
	\impl{}]{\includegraphics[width=\resultsize]{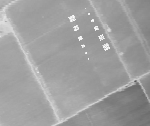}\label{fig:implcrop}}
	\hfil
	\subfloat[Ground 
	truth]{\includegraphics[width=\resultsize]{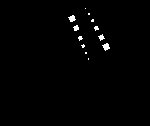}\label{fig:implgt}}
	\hfil
	\subfloat[\gls{rxd} ($t = 
	0.26$)]{\includegraphics[width=\resultsize]{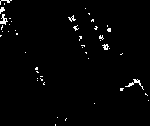}\label{fig:implrxt}}
	\hfil
	\subfloat[\gls{rsad}]{\includegraphics[width=\resultsize]{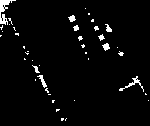}\label{fig:implrsadt}}
	\hfil
	\subfloat[\gls{wscf} ($t = 
	0.26$)]{\includegraphics[width=\resultsize]{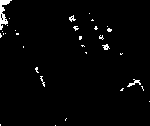}\label{fig:implwscft}}
	\hfil
	\subfloat[\gls{our} ($t = 
	0.22$)]{\includegraphics[width=\resultsize]{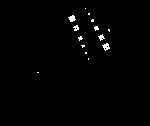}\label{fig:implladt}}
	\caption{\label{fig:hyperout}Hyperspectral test scenarios and 
			algorithm outputs. \gls{our} results have been obtained using 
			$\vLc$.}
\end{figure*}

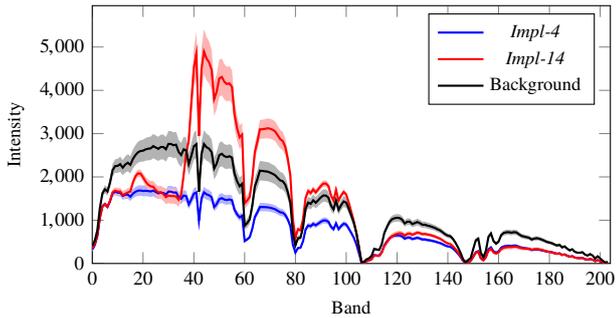
\begin{figure}
	\input{img/spectraplot.tex}
	\caption{\label{fig:spectraplot}Spectral characteristic curves 
		for different regions of the image. The line represents 
		the 
		mean intensity computed over all pixels in a region, while the shaded 
		area 
		represents the standard deviation.}
\end{figure}

\subsubsection{Dataset}

The dataset used in this study is composed of three hyperspectral 
scenes collected by the 224-band AVIRIS sensor. As a common 
practice~\cite{chang_anomaly_2002}, we discarded the 20 water absorption bands, 
\ie{} bands (108-112, 154-167, 224).
The first scene was collected over Salinas 
Valley, California, and is characterized by high spatial resolution (3.7-meter 
pixels). 
The area covered by this scene comprises 512 lines by 217 samples, and it 
includes vegetables, bare soils, and vineyard fields. A classification ground 
truth containing 16 classes is provided with this scene. A sample band of the 
image together with the classification ground truth is shown in 
\figref{fig:salinas}.
The other two scenes image two urban environments and come with anomaly 
detection ground truth, and both comprise 100 lines by 100 samples. We 
will refer to them as \urbA{} and \urbB{}. A sample band of these two scenes, 
together with their corresponding ground truth, is shown in 
\figref{fig:hyperout}.

\begin{figure*}
	\input{img/roc.tex}
	\caption{\Gls{roc} curves for the hyperspectral testing scenarios}
	\label{fig:roc}
\end{figure*}

To evaluate \gls{our}, we tested it on both real and synthetic anomalies.
For the Salinas scene, we cropped a $200 \times 150$ portion 
of the scene and manually segmented a construction which was visible in the 
cropped area: as the scene mostly contains fields of various kinds, this 
human-made construction was a good anomalous candidate. This setup, which we 
will call \real{}, is shown in \figref{fig:realcrop} together with its ground 
truth in \figref{fig:realgt}.

To obtain a synthetic anomaly, we used the target implant method 
\cite{stefanou_method_2009} on a different portion of the Salinas scene. The 
$150 \times 126$ binary mask image~$\vc{M}$
shown in \figref{fig:implgt} has been constructed by generating six squares 
having sides measuring from 1 to 6 pixels arranged in a line. The six squares 
have been then copied in reverse order and arranged in another line at close 
distance. The two lines have finally been rotated by an angle of 
approximatively~$\pi/6$. The 
pixels inside the squares have value of 1, while the rest of the pixels 
in~$\vc{M}$ have value 0.
Then, we cropped a region~$\vc{I}$ from the Salinas scene, having the same 
dimension as the mask. We used it to build the modified image~$\vc{I}'$ containing the 
implanted target as follows:
\begin{equation} \label{eq:impl}
\vc{I}'(i, j) = \vc{M}(i,j) \cdot \varPhi(k) + (1 - \vc{M}(i,j)) \cdot 
\vc{I}(i,j)\enspace,
\end{equation}
where~$\varPhi$ is a function that, given a parameter $k \in [1,16]$, returns a 
random pixel from the region of the Salinas scene having class~$k$ according to 
the classification ground truth shown in \figref{fig:salc}.
In the following discussion, for conciseness, we will limit the analysis to two 
synthetic setups with $k = 14$ and $k = 4$, respectively. The two 
representative values have been chosen since \gls{rxd} achieves the best 
performance on the former and the worst one on the latter. We will refer to 
them as \impl{} and \implB{}, respectively. A sample band from the \impl{} setup 
is shown in \figref{fig:implcrop}.

\figref{fig:spectraplot} shows the mean and standard deviation of the intensity of 
each band for the background, the anomaly region in \implB{} and \impl{}. As it 
can be noticed, the spectral characteristics of the anomaly in \implB{} are 
similar in shape to those of the background, although with reduced intensities. 
The anomaly in \impl{} presents a more different curve than the others, instead.

\subsubsection{Experimental results} \label{sec:res}

We are interested in evaluating the detection accuracy of \gls{our}
using the Laplacian model built over the partial correlation weights ($\vLn$) 
and the one built using Cauchy distance ($\vLc$). Also, we want to test both 
the spectral version of \gls{our} and its spatially aware variant \gls{our}-S.
The results will be compared with those yielded by classic \gls{rxd}, 
\gls{rsad}, and \gls{wscf}. 
We compare our results against those yielded by \gls{rxd}, given
its well known status as benchmark algorithm for anomaly detection. 
We want also to confirm with our experiments one of the known limitations of 
\gls{rxd} enunciated in \secref{sec:rx}, namely how the inclusion of spatial 
information in \gls{rxd} is detrimental to its performance, to demonstrate how 
our approach overcomes this limitation. Another well-known algorithm which aims 
at addressing this limitation is \gls{wscf}, and 
for this reason we selected it for evaluation as well. \Gls{wscf} requires a 
parameter~$\alpha$ to determine the amount of spatial information included in 
the metric. In this study, we set $\alpha = 0.2$, as suggested in the original 
work \cite{gaucel_whitening_2005}.
\Gls{rsad} requires to select: the initial number of randomly selected 
blocks $N$, which should be as small as possible but still large enough so that 
$4N>b$, where $b$ is the number of image bands; the number of random selections 
$L$; and the percentile $\alpha$. For these parameters, we chose the following 
values in our experiments: $N=80$, $L=40$, and $\alpha=0.001$.
We implemented our method as well as all three benchmark 
methods in MATLAB 2014b. All experiments were run on a laptop
equipped with an Intel\textsuperscript{\textregistered{}} Core\texttrademark{}
i7@2.20GHz CPU, a NVIDIA GT435M GeForce GPU and 8GB of RAM
\footnote{The hyperspectral datasets and all algorithm 
implementations used for the experiments presented in this work can be found 
at:\\ 
\href{https://github.com/fverdoja/LAD-Laplacian-Anomaly-Detector}{github.com/fverdoja/LAD-Laplacian-Anomaly-Detector}.}.

\figref{fig:hyperout} shows the visual results by \gls{our} 
($\vLc$) approach compared to the ones yielded by \gls{rxd}, \gls{rsad}, and 
\gls{wscf} on the all hyperspectral scenarios. It can be clearly noticed that the 
lower number of false positives \gls{our} is able to achieve against all other 
algorithms.

\figref{fig:roc} shows the \gls{roc} curves for the hyperspectral test cases, 
for all algorithms except \gls{rsad}. The approach by virtue of which 
\gls{rsad} selects which pixels are anomalous does not lend itself to be plotted 
in a \gls{roc} curve. The scale of the \gls{fpr} axis has been enhanced, as 
common in anomaly detection 
studies~\cite{baghbidi_improvement_2011,li_collaborative_2015,yuan_local_2014}, 
given the great difference in scale between the number of negative pixels and 
positive ones. It can be noticed how in all scenarios except \urbA{} 
our approach outperforms both \gls{rxd} and \gls{wscf}. On \urbA{}, all 
algorithms perform very similarly. Also, worth noticing is that 
the inclusion of spatial information yields limited improvements on the 
hyperspectral scenarios. When comparing results obtained by \gls{our} 
using~$\vLn$ or~$\vLc$, it can be noticed how performance is often very 
similar. This is a remarkable result, also considering that~$\vLc$ creates a 
model of the background without the need for matrix inversions, so it proves to 
be both quicker and equally precise.

To further compare performance yielded by the different approaches, we also use
the standard \gls{soi}~\cite{zou_statistical_2004}, also known as Dice 
similarity coefficient (DSC)~\cite{dice_measures_1945}, which can be 
computed as follows:
\begin{equation} \label{eq:soi}
SOI = \frac{2(A \cap B)}{A + B}\enspace,
\end{equation}
where~$A$ and~$B$ are two binary masks (\ie{} the ground truth or \gls{roi} and
the output of an automatic algorithm); the intersection operator is used
to indicate the number of pixels/voxels having value 1 in both masks, while the 
sum operator indicates the total number of pixels/voxels having value 1 in the 
two masks. \gls{soi} is also equivalent to the statistical $F_1$-score, which 
is the harmonic mean of precision and sensitivity, and is usually defined in 
terms of Type I and Type II errors as follows:
\begin{equation} \label{eq:f1}
F_1 = \frac{2 \cdot \mbox{true positive}}{2 \cdot \mbox{true positive} + 
	\mbox{false positive} + \mbox{false negative}}\,.
\end{equation}
The equality between \eqref{eq:soi} and \eqref{eq:f1} can be easily 
demonstrated considering that $A \cap B$ contains the true positive 
pixels/voxels and that if we consider that $A = (\mbox{true positive} + 
\mbox{false positive})$ and $B = (\mbox{true positive} + \mbox{false 
	negative})$, then also the denominator in \eqref{eq:soi} equals the one in 
\eqref{eq:f1}. Clearly, to compute the \gls{soi} metric one needs 
to select a threshold~$t$ to identify the anomaly subset~$B$. 
Many 
approaches~\cite{acito_cfar_2013,otsu_threshold_1979,zaidi_comparative_2012} 
have been proposed in the literature to deal with the problem of choosing the 
optimal threshold.
In this work, we select the value of~$t$ yielding the highest \gls{soi},
\ie{} striking the best balance between TPR and \gls{fpr} on the 
\gls{roc} curve in terms of \gls{soi}.
This choice allows us to compute a single-objective metric to compare the 
analyzed methods. 
Alternatively, we could also use the \gls{auc}, 
which measures the area under each \gls{roc} curve; we decided to avoid such 
metric since it has been recently criticized for being sensitive to 
noise~\cite{hanczar_small-sample_2010} and for other significant problems it 
shows in model comparison~\cite{hand_measuring_2009,lobo_auc:_2008}.

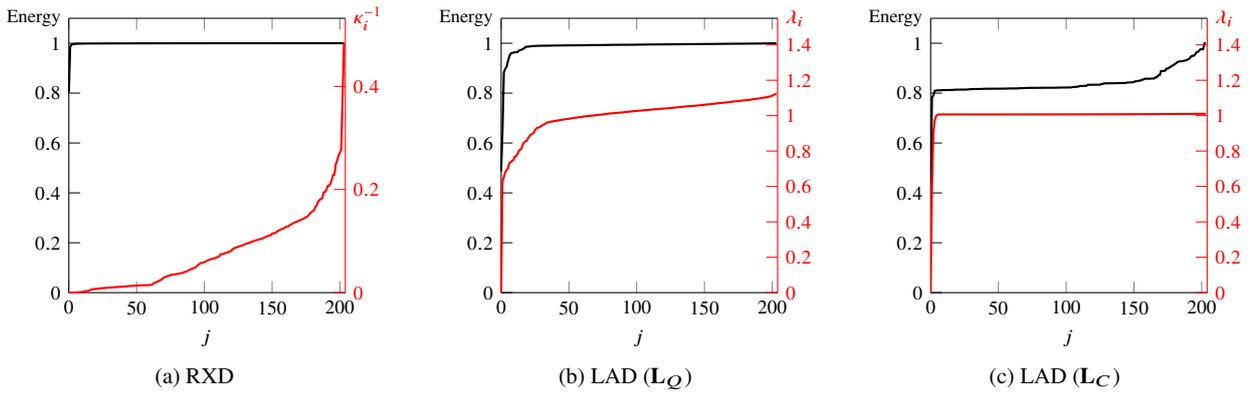
\begin{figure*}
	\input{img/energies.tex}
	\caption{\label{fig:energy}Energy and eigenvalue curves for the \impl{} 
		scenario}
\end{figure*}

\begin{table*}
	\caption{Experimental results in hyperspectral setup (\gls{soi}). For each 
	column, bold indicates the best result.}
	\label{tab:results}
	\centering
	\begin{tabular}{lcccccc}
		\toprule
		& \urbA{} & \urbB{} & \real{} & \impl{} & \implB{} & 
		Average \\
		\midrule
		\gls{rxd} & 0.508 & 0.649 & 0.685 & 0.445 & 0.045 & 0.466 \\
		\gls{rsad} & 0.078 & 0.310 & 0.042 & 0.450 & 0.022 & 0.180 \\
		\gls{wscf} & 0.489 & 0.623 & 0.708 & 0.391 & 0.103 & 0.463 \\
		\addlinespace
		\gls{our} ($\vLn$) & 0.606 & \textbf{0.791} & 0.806 & 0.941 & 0.525 & 
		\textbf{0.734} \\
		\gls{our}-S ($\vLn$) & 0.576 & 0.664 & \textbf{0.818} & 0.898 & 
		\textbf{0.540} & 0.699 \\
		\addlinespace
		\gls{our} ($\vLc$) & \textbf{0.614} & 0.782 & 0.754 & \textbf{0.954} & 
		0.514 & 0.724 \\
		\gls{our}-S ($\vLc$) & 0.467 & 0.721 & 0.697 & 0.919 & 0.409 & 0.643 \\
		\bottomrule
	\end{tabular} 
\end{table*}

\begin{table*}
	\caption{Experimental results after dimensionality reduction in 
		hyperspectral setup (\gls{soi}).  For each column, bold indicates the 
		best result.} 
	\label{tab:resultsp}
	\centering
	\begin{tabular}{lccccccc}
		\toprule
		& \urbA{} & \urbB{} & \real{} & \impl{} & \implB{} & 
		Average & Gain (\%)\\
		\midrule
		\gls{rxd}\textsuperscript{$p$} & \textbf{0.692} & 0.304 & 
		\textbf{0.930} & \textbf{0.965} & 0.355 & 0.649 & \textbf{+39.19} \\
		\addlinespace
		\gls{our}\textsuperscript{$p$} ($\vLn$) & 0.606 & \textbf{0.791} & 
		0.806 & 0.941 & 0.521 & \textbf{0.733} & -0.11 \\
		\gls{our}-S\textsuperscript{$p$} ($\vLn$) & 0.603 & 0.659 & 0.817 & 
		0.928 & \textbf{0.579} & 0.717 & +2.57 \\
		\addlinespace
		\gls{our}\textsuperscript{$p$} ($\vLc$) & 0.606 & 0.776 & 0.789 & 0.951 
		& 0.535 & 0.731 & +1.08 \\
		\gls{our}-S\textsuperscript{$p$} ($\vLc$) & 0.462 & 0.725 & 0.706 & 
		0.945 & 0.423 & 0.652 & +1.49 \\
		\bottomrule
	\end{tabular} 
\end{table*}

\tabref{tab:results} shows all \gls{soi} results of our tests. 
It can be noticed how all variants of our approach are able to outperform 
\gls{rxd}, \gls{rsad}, and \gls{wscf}.
These results are consistent with those presented by the \gls{roc} curves.

Finally, in \tabref{tab:resultsp} we show results of the de-noised version of 
both \gls{our} and \gls{rxd}, which we call \gls{our}\textsuperscript{$p$} and 
\gls{rxd}\textsuperscript{$p$}, respectively. 
In this case, the value of~$p$ has been chosen according to the cumulative 
energy as described in \secref{sec:rx}, setting $\psi = 0.99$. 
It can be  noticed how \gls{rxd} is able to 
gain the most from dimensionality reduction. These results can be 
explained considering the distribution of energy in the eigenspace 
decomposition. For the \impl{} scenario, in \figref{fig:energy} we show the 
cumulative energy distribution in the different eigenspaces  together 
with the corresponding eigenvalues~$\kappa_j^{-1}$ and~$\lambda_j$ (that are 
used to weigh the different contribution  in 
\eqref{eq:rxp} and \eqref{eq:ladp} respectively). It can be noticed that in the 
\gls{rxd} case (\figref{fig:energyrx}) energy is better compacted into few 
eigenspaces with respect to \gls{our} (\figref{fig:energyladq} and 
\figref{fig:energyladc}). At the same time,
it can be observed that the distribution of~$\kappa_j^{-1}$ in \gls{rxd} 
dramatically amplifies the last eigenspaces, \ie{} the noise components, 
according to \eqref{eq:rxp}.
On the contrary, this phenomenon does not affect \gls{our} since 
the distribution of eigenvalues~$\lambda_j$ is not peaked on the last 
eigenspaces. 
It follows that the effect of noise in \eqref{eq:ladp} is mitigated by 
construction and the benefit of dimensionality reduction is limited.
Indeed, it can be noted that results obtained 
by \gls{rxd} after dimensionality reduction are in line with those obtained by 
\gls{our} in its simple form. Being the eigendecomposition a costly operation, 
on a par with matrix inversion, the use of \gls{our} ($\vLc$), which does not 
require any  matrix inversion or eigendecomposition, might be preferable.

\begin{figure}
	\centering
	\includegraphics[width=\linewidth]{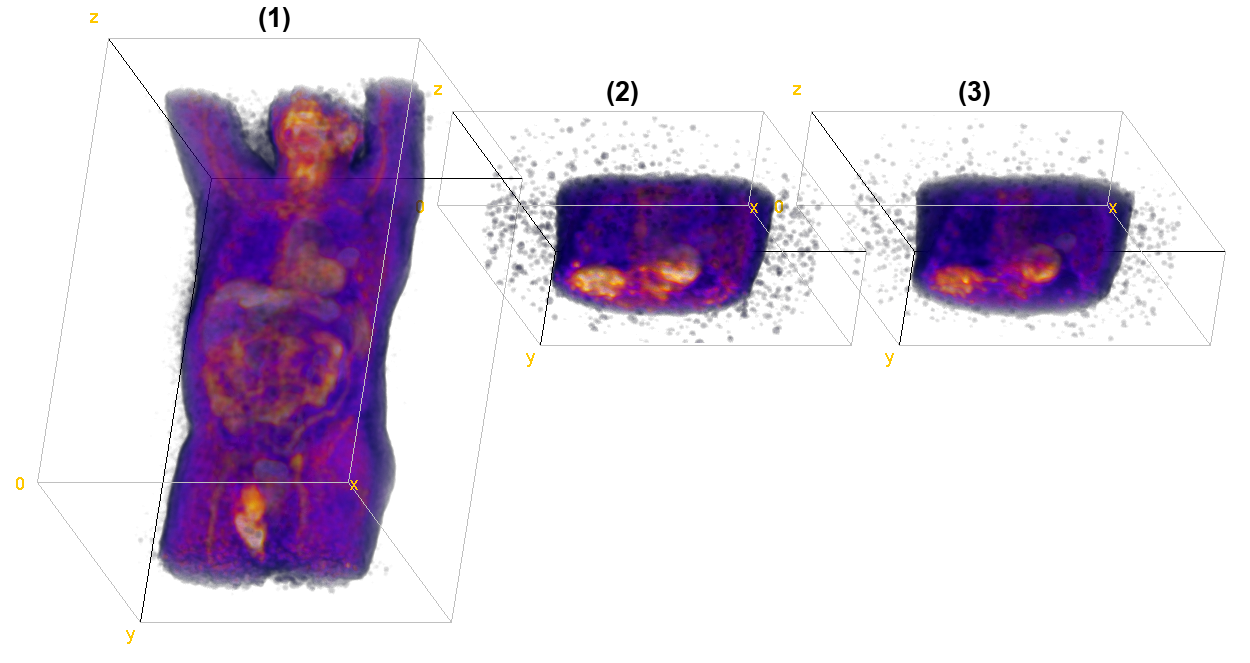}
	\caption{\label{fig:scans}The three FDG-PET images of one of the sample 
		patients; {\it (1)} is the early scan (\scanname{ES}, 
		144$\times$144$\times$213 px), {\it (2)} and {\it (3)} are constructed 
		integrating the delayed scan in 3-min time windows (\scanname{DS1} 
		and \scanname{DS2}, 144$\times$144$\times$45 px). Only the area 
		containing the tumor is acquired in the delayed scan. These images, 
		originally in grayscale, are here displayed using a {\it Fire} lookup 
		table.}
\end{figure}

\subsection{Application to 3D volumes: tumor segmentation in PET sequences} 
\label{sec:tumor}

\gls{pet} data are volumetric medical images that are usually employed to locate
the tumoral area for proper oncological treatment, \eg{} by means of 
radiotherapy.
From a \gls{pet} scan, one or more 3D images can be produced where the 
intensity of a voxel represents the local concentration of the tracer during the 
time window of the scan. 
In particular, \gls{fdg} is used to detect
tissue metabolic activity by virtue of the glucose uptake. 

During normal cell
replication, mutations in the DNA can occur and lead to the birth of cancer
cells. By their nature, these cells lack the ability to stop their
multiplication, raising cell density in their region and causing insufficient 
blood supply. The resulting deficiency in
oxygen (hypoxia) forces these cells to rely mostly on their anaerobic
metabolism, \ie{} glycolysis~\cite{halperin_principles_2008}. For this 
reason, glycolysis is an excellent marker for detecting cancer cells; \gls{fdg} 
---in which the tracer's concentration indicates the glucose uptake in the 
imaged area---turns out to be a suitable tool for recognizing tumors, 
metastases, and lymph nodes all at once~\cite{garber_energy_2004}.
It follows that proper segmentation of tumors in medical images is crucial as 
oncological treatment plans rely on precise information on the tumoral region 
to be effective~\cite{halperin_principles_2008}.
Manual segmentation by medical staff has been proven to be subjective,
inaccurate, and time-consuming~\cite{wong_segmentation_2002}; for this reason,
the need for automatic methods for tumor region segmentation is on the rise.
\Gls{pet} images carry information about cells metabolism and are therefore
suitable for this task; however, \gls{pet} segmentation is still an open
problem mainly because of limited image resolution and strong presence of
acquisition noise~\cite{zaidi_comparative_2012}.

\begin{table}
	\caption{Experimental results in \tumor{} setup (\gls{soi}). Bold indicates 
	the best result.}
	\label{tab:resultsT}
	\centering
	\begin{tabular}{lc}
		\toprule
		& Average \\
		\midrule
		\gls{rxd} & 0.570 \\
		\addlinespace
		\gls{our} ($\vLn$) & 0.362 \\
		\gls{our}-S ($\vLn$) & \textbf{0.592} \\
		\addlinespace
		\gls{our} ($\vLc$) & 0.427 \\
		\gls{our}-S ($\vLc$) & 0.560 \\
		\bottomrule
	\end{tabular} 
\end{table}

In~\cite{bracco_automatic_2016,verdoja_global_2016,verdoja_automatic_2014}, we 
successfully explored the use of \gls{rxd} to identify the anomalous behavior 
of cancer cells over time in sequences of three \gls{fdg} images acquired over 
a time span of one hour. A quick visual overview of this setup is shown in 
\figref{fig:scans}. The idea behind the use of \gls{rxd} in this scenario arises
from the fact that cancer cells tend to acquire glucose differently than normal 
cells, given their peculiar reliance on anaerobic metabolism. For this reason, 
when considering the values a voxel assumes over time, cancer's 
anomalous glucose uptake can be successfully spotted using anomaly detection 
techniques, where the usual role of spectral bands is taken by three PET images
acquired over time.

To do this, we build a 4D matrix~$\vc{I}$, having the 
three spatial dimensions as the first three dimensions and the time as the fourth 
dimension. Being acquired at different times, with the subject assuming 
slightly different positions, it is worth recalling that the images need to be 
aligned using registration algorithms as detailed 
in~\cite{verdoja_automatic_2014}. The resulting matrix~$\vc{I}$ will then have 
size $144 \times 144 \times 45 \times 3$. Then, for a 
generic voxel, identified by its spatial coordinates, we define the vector 
$\x{} = [x_1 x_2 x_3]^T$ as the vector containing that voxel's intensities over 
time. In other words, RXD can be employed in this case if  time  takes the role
 of the spectral dimension.

\begin{figure*}
	\input{img/rocT.tex}
	\caption{\gls{roc} curves for all patients in the \tumor{} testing 
			scenario}
	\label{fig:rocT}
\end{figure*}
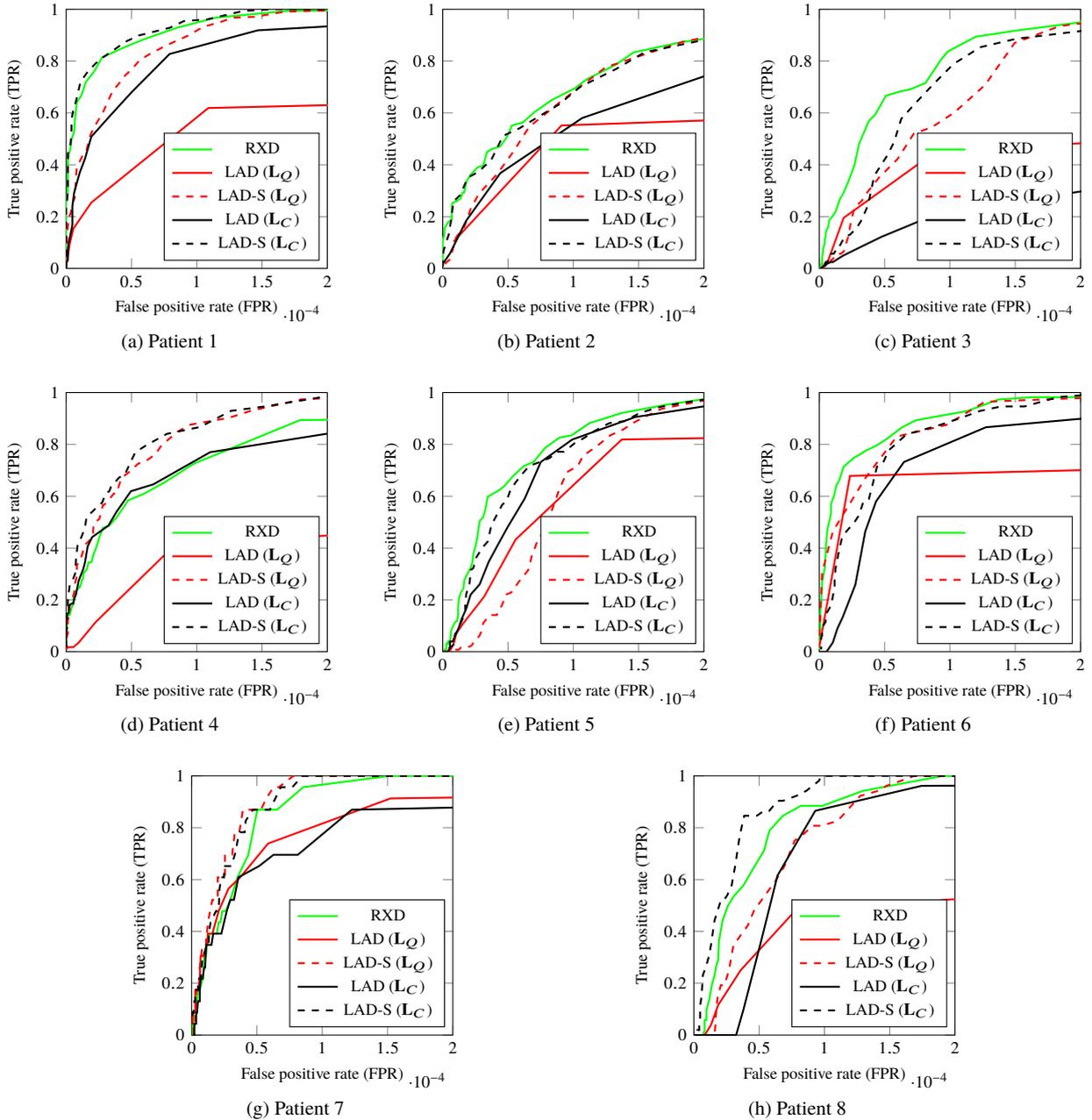

\subsubsection{Experimental results}

In this study, we used a dataset comprising eight patients, that has been made 
available by the \gls{ircc} for research purposes. All the acquisitions have 
been made using a {\it Philips Gemini TF PET/CT}. To this end, we acknowledge 
the precious aid of nuclear medicine physicians who have manually segmented the 
\glspl{roi} on the \gls{pet} images, setting up the ground truth for evaluating 
the performance yielded by the proposed tools. We will refer to this setup as 
\tumor{}.

Also in this scenario, we are interested in evaluating the detection accuracy 
of \gls{our} using both Laplacian models,~$\vLn$ and~$\vLc$, and compare our 
results with those yielded by classic \gls{rxd}. We cannot compare with 
\gls{wscf} in this domain as its extension to 3D has not been proposed, and 
therefore the choice of the parameter~$\alpha$ is non-trivial.
A thing to notice regarding this setup is that we are dealing with voxels and 3D
volumes. For this reason, in \gls{our}-S we will use 6-connectivity, which is
the extension of 2D 4-connectivity to 3D space.

To compare performance yielded by the different approaches, we use
\gls{soi} as presented in \eqref{eq:soi}. Once again, in this study we 
selected the value of~$t$ yielding the highest \gls{soi}.

\figref{fig:rocT} shows the \gls{roc} curves for all the eight patients in the 
\tumor{} dataset, while \tabref{tab:resultsT} shows the average \gls{soi} 
results of our tests over the patient dataset. The inclusion of spatial 
information in the graph improves the \gls{soi} metric. In this scenario, we do 
not present results after dimensionality reduction 
because the spectral dimensions were already very few. 
Also, in this scenario the use of \gls{our} is able to obtain performance 
similar when not better than \gls{rxd} in all its variances.

%% file: img/spectraplot.tex
\centering
\usepgfplotslibrary{fillbetween}
\begin{tikzpicture}
	\begin{axis}[
		xlabel={Band},
		ylabel={Intensity},
		label style={font=\scriptsize}, tick label style={font=\scriptsize}, 
		width=\linewidth, height=5cm,
		xtick distance=20,ytick distance=1000,
		xmin=0, xmax=204, ymin=0,
		legend style={font=\scriptsize}, legend pos=north east,
		legend entries={
			\implB{},
			\impl{},
			Background
		}
	]
	
%
	
	\addplot[name path=s4a,opacity=0,forget plot] table[x expr=\coordindex, 
	y=s4a] 
	{img/spectra.txt};
	\addplot[name path=4,thick,blue] table[x expr=\coordindex, y=m4] 
	{img/spectra.txt};
	\addplot[name path=s4b,opacity=0,forget plot] table[x expr=\coordindex, 
	y=s4b] 
	{img/spectra.txt};
	
	\addplot[blue!30,forget plot] fill between[of=s4a and s4b];
		
	\addplot[name path=s14a,opacity=0,forget plot] table[x expr=\coordindex, 
	y=s14a] 
	{img/spectra.txt};
	\addplot[name path=14,thick,red] table[x expr=\coordindex, y=m14] 
	{img/spectra.txt};
	\addplot[name path=s14b,opacity=0,forget plot] table[x expr=\coordindex, 
	y=s14b] 
	{img/spectra.txt};
	
	\addplot[red!30,forget plot] fill between[of=s14a and s14b];
	
	\addplot[name path=sBa,opacity=0,forget plot] table[x expr=\coordindex, 
	y=sBa] 
	{img/spectra.txt};
	\addplot[name path=B,thick,black] table[x expr=\coordindex, y=mB] 
	{img/spectra.txt};
	\addplot[name path=sBb,opacity=0,forget plot] table[x expr=\coordindex, 
	y=sBb] 
	{img/spectra.txt};
	
	\addplot[black!30,forget plot] fill between[of=sBa and sBb];
	
	\end{axis}
\end{tikzpicture}

%% file: img/roc.tex
\centering
   \subfloat[\urbA{}]{
      \begin{tikzpicture}
      \begin{axis}[
      xlabel={False positive rate (FPR)},
      ylabel={True positive rate (TPR)},
      xmin=0, xmax=0.06,
      ymin=0, ymax=1,
      label style={font=\scriptsize}, tick label style={font=\scriptsize}, 
      legend style={font=\scriptsize}, width=\rocsize, height=\rocsize,
      legend pos=south east,
      legend entries={
         \gls{rxd},
         \gls{wscf},
         \gls{our} ($\vLn$),
         \gls{our}-S ($\vLn$),
         \gls{our} ($\vLc$),
         \gls{our}-S ($\vLc$),
      }
      ]
      \addplot[thick,green] table[x=Xrx, y=Yrx] {img/rocU1.txt};
      \addplot[thick,green,dashed] table[x=Xwscf, y=Ywscf] {img/rocU1.txt};
      \addplot[thick,red] table[x=Xq, y=Yq] {img/rocU1.txt};
      \addplot[thick,red,dashed] table[x=Xq-s, y=Yq-s] {img/rocU1.txt};
      \addplot[thick,black] table[x=Xc, y=Yc] {img/rocU1.txt};
      \addplot[thick,black,dashed] table[x=Xc-s, y=Yc-s] {img/rocU1.txt};
      \end{axis}
      \end{tikzpicture}\label{fig:rocurb1}
   }
   \hfil
   \subfloat[\urbB{}]{
   	\begin{tikzpicture}
   	\begin{axis}[
   	xlabel={False positive rate (FPR)},
   	ylabel={True positive rate (TPR)},
   	xmin=0, xmax=0.06,
   	ymin=0, ymax=1,
   	label style={font=\scriptsize}, tick label style={font=\scriptsize}, 
   	legend style={font=\scriptsize}, width=\rocsize, height=\rocsize,
   	legend pos=south east,
   	legend entries={
   		\gls{rxd},
   		\gls{wscf},
   		\gls{our} ($\vLn$),
   		\gls{our}-S ($\vLn$),
   		\gls{our} ($\vLc$),
   		\gls{our}-S ($\vLc$),
   	}
   	]
   	\addplot[thick,green] table[x=Xrx, y=Yrx] {img/rocU2.txt};
   	\addplot[thick,green,dashed] table[x=Xwscf, y=Ywscf] {img/rocU2.txt};
   	\addplot[thick,red] table[x=Xq, y=Yq] {img/rocU2.txt};
   	\addplot[thick,red,dashed] table[x=Xq-s, y=Yq-s] {img/rocU2.txt};
   	\addplot[thick,black] table[x=Xc, y=Yc] {img/rocU2.txt};
   	\addplot[thick,black,dashed] table[x=Xc-s, y=Yc-s] {img/rocU2.txt};
   	\end{axis}
   	\end{tikzpicture}\label{fig:rocurb2}
   }
   \hfil
   \subfloat[\real{}]{
   	\begin{tikzpicture}
   	\begin{axis}[
   	xlabel={False positive rate (FPR)},
   	ylabel={True positive rate (TPR)},
   	xmin=0, xmax=0.005,
   	ymin=0, ymax=1,
   	label style={font=\scriptsize}, tick label style={font=\scriptsize}, 
   	legend style={font=\scriptsize}, width=\rocsize, height=\rocsize,
   	legend pos=south east,
   	legend entries={
   		\gls{rxd},
   		\gls{wscf},
   		\gls{our} ($\vLn$),
   		\gls{our}-S ($\vLn$),
   		\gls{our} ($\vLc$),
   		\gls{our}-S ($\vLc$),
   	}
   	]
   	\addplot[thick,green] table[x=Xrx, y=Yrx] {img/rocR.txt};
   	\addplot[thick,green,dashed] table[x=xR, y=yR] {img/rocWSCF.txt};
   	\addplot[thick,red] table[x=Xq, y=Yq] {img/rocR.txt};
   	\addplot[thick,red,dashed] table[x=Xq-s, y=Yq-s] {img/rocR.txt};
   	\addplot[thick,black] table[x=Xc, y=Yc] {img/rocR.txt};
   	\addplot[thick,black,dashed] table[x=Xc-s, y=Yc-s] {img/rocR.txt};
   	\end{axis}
   	\end{tikzpicture}\label{fig:rocreal}
   }
   \hfil
   \subfloat[\implB{}]{
      \begin{tikzpicture}
      \begin{axis}[
      xlabel={False positive rate (FPR)},
      ylabel={True positive rate (TPR)},
      xmin=0, xmax=0.4,
      ymin=0, ymax=1,
      label style={font=\scriptsize}, tick label style={font=\scriptsize}, 
      legend style={font=\scriptsize}, width=\rocsize, height=\rocsize,
      legend pos=south east,
      legend entries={
         \gls{rxd},
         \gls{wscf},
         \gls{our} ($\vLn$),
         \gls{our}-S ($\vLn$),
         \gls{our} ($\vLc$),
         \gls{our}-S ($\vLc$),
      }
      ]
      \addplot[thick,green] table[x=Xrx, y=Yrx] {img/roc4.txt};
      \addplot[thick,green,dashed] table[x=x4, y=y4] {img/rocWSCF.txt};
      \addplot[thick,red] table[x=Xq, y=Yq] {img/roc4.txt};
      \addplot[thick,red,dashed] table[x=Xq-s, y=Yq-s] {img/roc4.txt};
      \addplot[thick,black] table[x=Xc, y=Yc] {img/roc4.txt};
      \addplot[thick,black,dashed] table[x=Xc-s, y=Yc-s] {img/roc4.txt};
      \end{axis}
      \end{tikzpicture}\label{fig:roc14}
   }
   \hfil
   \subfloat[\impl{}]{
      \begin{tikzpicture}
      \begin{axis}[
      xlabel={False positive rate (FPR)},
      ylabel={True positive rate (TPR)},
      xmin=0, xmax=0.02,
      ymin=0, ymax=1,
      label style={font=\scriptsize}, tick label style={font=\scriptsize}, 
      legend style={font=\scriptsize}, width=\rocsize, height=\rocsize,
      legend pos=south east,
      legend entries={
         \gls{rxd},
         \gls{wscf},
         \gls{our} ($\vLn$),
         \gls{our}-S ($\vLn$),
         \gls{our} ($\vLc$),
         \gls{our}-S ($\vLc$),
      }
      ]
      \addplot[thick,green] table[x=Xrx, y=Yrx] {img/roc14.txt};
      \addplot[thick,green,dashed] table[x=x14, y=y14] {img/rocWSCF.txt};
      \addplot[thick,red] table[x=Xq, y=Yq] {img/roc14.txt};
      \addplot[thick,red,dashed] table[x=Xq-s, y=Yq-s] {img/roc14.txt};
      \addplot[thick,black] table[x=Xc, y=Yc] {img/roc14.txt};
      \addplot[thick,black,dashed] table[x=Xc-s, y=Yc-s] {img/roc14.txt};
      \end{axis}
      \end{tikzpicture}\label{fig:roc4}
   }

%% file: img/energies.tex
\centering
\subfloat[\gls{rxd}]{
  \begin{tikzpicture}
  \pgfplotsset{
     label style={font=\scriptsize}, tick label style={font=\scriptsize}, 
     width=\energysize, height=\energysize,
     xtick distance=50,ytick distance=0.2,
     xmin=0, xmax=204,
     y axis style/.style={
        yticklabel style=#1,
        ylabel style=#1,
        y axis line style=#1,
        ytick style=#1
     }
  }

  \begin{axis}[
  axis y line*=left,
  y axis style=black,
  ylabel style={at={(0,1)},rotate=-90},
  ymin=0,
  xlabel=$j$,
  ylabel=Energy
  ]
  \addplot[thick,black] table[x expr=\coordindex, y=rxCum] 
  {img/energy14.txt};
  \end{axis}
  \begin{axis}[
  axis y line*=right,
  axis x line=none,
  ylabel style={at={(1,1)},rotate=-90},
  ymin=0,
  ylabel=$\kappa_i^{-1}$,
  y axis style=red
  ]
  \addplot[thick,red] table[x expr=\coordindex, y=rxLambda] 
  {img/energy14.txt};
  \end{axis}
  \end{tikzpicture}\label{fig:energyrx}
}
\hfil
\subfloat[\gls{our} ($\vLn$)]{
  \begin{tikzpicture}
  \pgfplotsset{
     label style={font=\scriptsize}, tick label style={font=\scriptsize}, 
     width=\energysize, height=\energysize,
     xtick distance=50,ytick distance=0.2,
     xmin=0, xmax=204,
     y axis style/.style={
        yticklabel style=#1,
        ylabel style=#1,
        y axis line style=#1,
        ytick style=#1
     }
  }
  
  \begin{axis}[
  axis y line*=left,
  y axis style=black,
  ylabel style={at={(0,1)},rotate=-90},
  ymin=0,
  xlabel=$j$,
  ylabel=Energy
  ]
  \addplot[thick,black] table[x expr=\coordindex, y=qCum] 
  {img/energy14.txt};
  \end{axis}
  \begin{axis}[
  axis y line*=right,
  axis x line=none,
  ylabel style={at={(1,1)},rotate=-90},
  ymin=0,ymax=1.55,
  ylabel=$\lambda_i$,
  y axis style=red
  ]
  \addplot[thick,red] table[x expr=\coordindex, y=qLambda] 
  {img/energy14.txt};
  \end{axis}
  \end{tikzpicture}\label{fig:energyladq}
}
\hfil
\subfloat[\gls{our} ($\vLc$)]{
  \begin{tikzpicture}
  \pgfplotsset{
     label style={font=\scriptsize}, tick label style={font=\scriptsize}, 
     width=\energysize, height=\energysize,
     xtick distance=50,ytick distance=0.2,
     xmin=0, xmax=204,
     y axis style/.style={
        yticklabel style=#1,
        ylabel style=#1,
        y axis line style=#1,
        ytick style=#1
     }
  }
  
  \begin{axis}[
  axis y line*=left,
  y axis style=black,
  ylabel style={at={(0,1)},rotate=-90},
  ymin=0,
  xlabel=$j$,
  ylabel=Energy
  ]
  \addplot[thick,black] table[x expr=\coordindex, y=cCum] 
  {img/energy14.txt};
  \end{axis}
  \begin{axis}[
  axis y line*=right,
  axis x line=none,
  ylabel style={at={(1,1)},rotate=-90},
  ymin=0,ymax=1.55,
  ylabel=$\lambda_i$,
  y axis style=red
  ]
  \addplot[thick,red] table[x expr=\coordindex, y=cLambda] 
  {img/energy14.txt};
  \end{axis}
  \end{tikzpicture}\label{fig:energyladc}
}

%% file: img/rocT.tex
\centering
\subfloat[Patient 1]{
	\begin{tikzpicture}
	\begin{axis}[
	xlabel={False positive rate (FPR)},
	ylabel={True positive rate (TPR)},
	xmin=0, xmax=0.0002,
	ymin=0, ymax=1,
	label style={font=\scriptsize}, 
	tick label style={font=\scriptsize}, 
	width=\rocsizeT, height=\rocsizeT, 
	legend style={font=\scriptsize}, legend pos=south east,
	legend entries={
		\gls{rxd},
		\gls{our} ($\vLn$),
		\gls{our}-S ($\vLn$),
		\gls{our} ($\vLc$),
		\gls{our}-S ($\vLc$)
	}
	]
	\addplot[thick,green] table[x=Xrx, y=Yrx] {img/rocT1.txt};
	\addplot[thick,red] table[x=Xq, y=Yq] {img/rocT1.txt};
	\addplot[thick,red,dashed] table[x=Xq-s, y=Yq-s] {img/rocT1.txt};
	\addplot[thick,black] table[x=Xc, y=Yc] {img/rocT1.txt};
	\addplot[thick,black,dashed] table[x=Xc-s, y=Yc-s] {img/rocT1.txt};
	\end{axis}
	\end{tikzpicture}\label{fig:roct1}
}
\hfil
\subfloat[Patient 2]{
	\begin{tikzpicture}
	\begin{axis}[
	xlabel={False positive rate (FPR)},
	ylabel={True positive rate (TPR)},
	xmin=0, xmax=0.0002,
	ymin=0, ymax=1,
	label style={font=\scriptsize}, 
	tick label style={font=\scriptsize}, 
	width=\rocsizeT, height=\rocsizeT, 
	legend style={font=\scriptsize}, legend pos=south east,
	legend entries={
		\gls{rxd},
		\gls{our} ($\vLn$),
		\gls{our}-S ($\vLn$),
		\gls{our} ($\vLc$),
		\gls{our}-S ($\vLc$)
	}
	]
	\addplot[thick,green] table[x=Xrx, y=Yrx] {img/rocT2.txt};
	\addplot[thick,red] table[x=Xq, y=Yq] {img/rocT2.txt};
	\addplot[thick,red,dashed] table[x=Xq-s, y=Yq-s] {img/rocT2.txt};
	\addplot[thick,black] table[x=Xc, y=Yc] {img/rocT2.txt};
	\addplot[thick,black,dashed] table[x=Xc-s, y=Yc-s] {img/rocT2.txt};
	\end{axis}
	\end{tikzpicture}\label{fig:roct2}
}
\hfil
\subfloat[Patient 3]{
	\begin{tikzpicture}
	\begin{axis}[
	xlabel={False positive rate (FPR)},
	ylabel={True positive rate (TPR)},
	xmin=0, xmax=0.0002,
	ymin=0, ymax=1,
	label style={font=\scriptsize}, 
	tick label style={font=\scriptsize}, 
	width=\rocsizeT, height=\rocsizeT, 
	legend style={font=\scriptsize}, legend pos=south east,
	legend entries={
		\gls{rxd},
		\gls{our} ($\vLn$),
		\gls{our}-S ($\vLn$),
		\gls{our} ($\vLc$),
		\gls{our}-S ($\vLc$)
	}
	]
	\addplot[thick,green] table[x=Xrx, y=Yrx] {img/rocT3.txt};
	\addplot[thick,red] table[x=Xq, y=Yq] {img/rocT3.txt};
	\addplot[thick,red,dashed] table[x=Xq-s, y=Yq-s] {img/rocT3.txt};
	\addplot[thick,black] table[x=Xc, y=Yc] {img/rocT3.txt};
	\addplot[thick,black,dashed] table[x=Xc-s, y=Yc-s] {img/rocT3.txt};
	\end{axis}
	\end{tikzpicture}\label{fig:roct3}
}
\hfil
\subfloat[Patient 4]{
	\begin{tikzpicture}
	\begin{axis}[
	xlabel={False positive rate (FPR)},
	ylabel={True positive rate (TPR)},
	xmin=0, xmax=0.0002,
	ymin=0, ymax=1,
	label style={font=\scriptsize}, 
	tick label style={font=\scriptsize}, 
	width=\rocsizeT, height=\rocsizeT, 
	legend style={font=\scriptsize}, legend pos=south east,
	legend entries={
		\gls{rxd},
		\gls{our} ($\vLn$),
		\gls{our}-S ($\vLn$),
		\gls{our} ($\vLc$),
		\gls{our}-S ($\vLc$)
	}
	]
	\addplot[thick,green] table[x=Xrx, y=Yrx] {img/rocT4.txt};
	\addplot[thick,red] table[x=Xq, y=Yq] {img/rocT4.txt};
	\addplot[thick,red,dashed] table[x=Xq-s, y=Yq-s] {img/rocT4.txt};
	\addplot[thick,black] table[x=Xc, y=Yc] {img/rocT4.txt};
	\addplot[thick,black,dashed] table[x=Xc-s, y=Yc-s] {img/rocT4.txt};
	\end{axis}
	\end{tikzpicture}\label{fig:roct4}
}
\hfil
\subfloat[Patient 5]{
	\begin{tikzpicture}
	\begin{axis}[
	xlabel={False positive rate (FPR)},
	ylabel={True positive rate (TPR)},
	xmin=0, xmax=0.0002,
	ymin=0, ymax=1,
	label style={font=\scriptsize}, 
	tick label style={font=\scriptsize}, 
	width=\rocsizeT, height=\rocsizeT, 
	legend style={font=\scriptsize}, legend pos=south east,
	legend entries={
		\gls{rxd},
		\gls{our} ($\vLn$),
		\gls{our}-S ($\vLn$),
		\gls{our} ($\vLc$),
		\gls{our}-S ($\vLc$)
	}
	]
	\addplot[thick,green] table[x=Xrx, y=Yrx] {img/rocT5.txt};
	\addplot[thick,red] table[x=Xq, y=Yq] {img/rocT5.txt};
	\addplot[thick,red,dashed] table[x=Xq-s, y=Yq-s] {img/rocT5.txt};
	\addplot[thick,black] table[x=Xc, y=Yc] {img/rocT5.txt};
	\addplot[thick,black,dashed] table[x=Xc-s, y=Yc-s] {img/rocT5.txt};
	\end{axis}
	\end{tikzpicture}\label{fig:roct5}
}
\hfil
\subfloat[Patient 6]{
	\begin{tikzpicture}
	\begin{axis}[
	xlabel={False positive rate (FPR)},
	ylabel={True positive rate (TPR)},
	xmin=0, xmax=0.0002,
	ymin=0, ymax=1,
	label style={font=\scriptsize}, 
	tick label style={font=\scriptsize}, 
	width=\rocsizeT, height=\rocsizeT, 
	legend style={font=\scriptsize}, legend pos=south east,
	legend entries={
		\gls{rxd},
		\gls{our} ($\vLn$),
		\gls{our}-S ($\vLn$),
		\gls{our} ($\vLc$),
		\gls{our}-S ($\vLc$)
	}
	]
	\addplot[thick,green] table[x=Xrx, y=Yrx] {img/rocT6.txt};
	\addplot[thick,red] table[x=Xq, y=Yq] {img/rocT6.txt};
	\addplot[thick,red,dashed] table[x=Xq-s, y=Yq-s] {img/rocT6.txt};
	\addplot[thick,black] table[x=Xc, y=Yc] {img/rocT6.txt};
	\addplot[thick,black,dashed] table[x=Xc-s, y=Yc-s] {img/rocT6.txt};
	\end{axis}
	\end{tikzpicture}\label{fig:roct6}
}
\hfil
\subfloat[Patient 7]{
	\begin{tikzpicture}
	\begin{axis}[
	xlabel={False positive rate (FPR)},
	ylabel={True positive rate (TPR)},
	xmin=0, xmax=0.0002,
	ymin=0, ymax=1,
	label style={font=\scriptsize}, 
	tick label style={font=\scriptsize}, 
	width=\rocsizeT, height=\rocsizeT, 
	legend style={font=\scriptsize}, legend pos=south east,
	legend entries={
		\gls{rxd},
		\gls{our} ($\vLn$),
		\gls{our}-S ($\vLn$),
		\gls{our} ($\vLc$),
		\gls{our}-S ($\vLc$)
	}
	]
	\addplot[thick,green] table[x=Xrx, y=Yrx] {img/rocT7.txt};
	\addplot[thick,red] table[x=Xq, y=Yq] {img/rocT7.txt};
	\addplot[thick,red,dashed] table[x=Xq-s, y=Yq-s] {img/rocT7.txt};
	\addplot[thick,black] table[x=Xc, y=Yc] {img/rocT7.txt};
	\addplot[thick,black,dashed] table[x=Xc-s, y=Yc-s] {img/rocT7.txt};
	\end{axis}
	\end{tikzpicture}\label{fig:roct7}
}
\hfil
\subfloat[Patient 8]{
	\begin{tikzpicture}
	\begin{axis}[
	xlabel={False positive rate (FPR)},
	ylabel={True positive rate (TPR)},
	xmin=0, xmax=0.0002,
	ymin=0, ymax=1,
	label style={font=\scriptsize}, 
	tick label style={font=\scriptsize}, 
	width=\rocsizeT, height=\rocsizeT, 
	legend style={font=\scriptsize}, legend pos=south east,
	legend entries={
		\gls{rxd},
		\gls{our} ($\vLn$),
		\gls{our}-S ($\vLn$),
		\gls{our} ($\vLc$),
		\gls{our}-S ($\vLc$)
	}
	]
	\addplot[thick,green] table[x=Xrx, y=Yrx] {img/rocT8.txt};
	\addplot[thick,red] table[x=Xq, y=Yq] {img/rocT8.txt};
	\addplot[thick,red,dashed] table[x=Xq-s, y=Yq-s] {img/rocT8.txt};
	\addplot[thick,black] table[x=Xc, y=Yc] {img/rocT8.txt};
	\addplot[thick,black,dashed] table[x=Xc-s, y=Yc-s] {img/rocT8.txt};
	\end{axis}
	\end{tikzpicture}\label{fig:roct8}
}

%% file: sections/6_disc.tex
\section{Discussion} \label{sec:disc}

In the previous section, we conducted experiments in hyperspectral and medical 
domain. \Gls{rxd}'s limitations detailed in \secref{sec:back} can be noticed in 
many of the presented experiments. In particular, the high number of false 
negative can be easily noticed in \figref{fig:hyperout}, while the poor 
performance of \gls{rxd}, \gls{rsad}, and \gls{wscf} for the \implB{} 
scenario can be imputed to the fact that in that case the
anomaly has a very similar covariance matrix to the background as 
shown in \figref{fig:spectraplot}; this makes very difficult for 
covariance-based methods to find an acceptable solution. 

The results obtained by \gls{rsad} have been particularly surprising. 
The algorithm has been able to achieve results inline or even better than the 
other two covariance-based approaches in a couple of scenarios, while obtaining 
very poor performance in the others due to very high \gls{fpr}. We believe this 
behavior is caused by the assumption made by \gls{rsad} while marking pixels as 
anomalous that the Mahalanobis distance follows a $\chi$ distribution. In the 
scenarios used in this study, we observed that that was rarely the case. When 
this assumption does not hold, the decision criterion used by \gls{rsad} is 
probably not sufficient.

The proposed technique was able to outperform state-of-the-art techniques
in all scenarios, proving how the flexibility of a graph model can actually 
enable better and more robust background estimation as well as successful 
inclusion of spatial information.

Spatially aware variants of the proposed techniques were able to 
achieve better performance in the \tumor{} scenarios, while failing at 
improving the performance of the spectral-only variants in the hyperspectral ones. 
The benefit of including spatial information is more noticeable in the medical 
scenario because in that case the spectral dimension is reduced to only three 
bands, representing three different acquisitions in time, as opposed to the 204 
spectral bands of the hyperspectral images. Also, we used a uniform correlation 
as model for the spatial weights; a more refined model might be more suited to 
better capture the spatial dynamics of remote sensing, while the one 
used might just be more fitting for medical imaging.

When comparing results obtained by \gls{our} using~$\vLn$ or~$\vLc$, it can be 
noticed how performance is often very similar on hyperspectral images, while 
in \tumor{}~$\vLc$ is able to obtain consistently better results.  
This behavior is clearly due to the fact that~$\vLn$ depends on pairwise 
correlation estimates that are particularly critical in the \tumor{} case, where 
the 3D volumes are characterized by 
poor spatiotemporal resolution. In this case, the use of graph prior based 
on~$\vLc$ turns out to be more robust.
An analysis of the \glspl{roc} validated this observation 
even further: for the hyperspectral case, the \gls{roc} curves for \gls{our}
using~$\vLn$ or~$\vLc$ behave very similarly in both cases, indicating that the
two weight functions are able to capture the same aspects of the data, while in
the \tumor{} case, the two \gls{roc} curves have a more varied behavior.

All these tests confirm that the use of our approach is preferable to 
\gls{rxd}, \gls{rsad}, and \gls{wscf} and that Laplacian estimated using the Cauchy 
distance is able to perform as well as the one estimated using partial 
correlation. Once again, this is remarkable as the former does not require any 
matrix inversion, while the latter does.

%% file: sections/7_concl.tex
\section{Conclusions} \label{sec:concl}
We present Laplacian anomaly detector, a graph-based algorithm aiming at 
detecting targets by virtue of a Laplacian model of the image background. Two 
different approaches to the graph construction are proposed. When comparing to 
\gls{rxd}, \gls{rsad}, and \gls{wscf}, one of the main advantages of our 
technique is its ability to model the image content without the need for matrix 
inversions.
Both visual inspection and objective results show how the proposed approach is 
able to outperform the other benchmark methods. Future direction might be 
devoted to evaluate \gls{our} ability to detect anomalies on generic non-image 
graphs.